\newcommand{\figwidth}{\linewidth}
\documentclass[%
 reprint,
superscriptaddress,
 amsmath,amssymb,
 aps,
]{revtex4-1}
\newcommand{\specificthanks}[1]{\@fnsymbol{#1}}

\usepackage{graphicx}
\usepackage{dcolumn}
\usepackage{bm}
\usepackage[colorlinks]{hyperref}
\hypersetup{urlcolor=blue,linkcolor=blue,citecolor=blue}

\usepackage{siunitx}

\preprint{APS/123-QED}
\begin{document}
\title{Reduced volume and reflection for bright optical tweezers with radial Laguerre-Gauss beams}
\author{J.-B. B\'eguin}\affiliation{Norman Bridge Laboratory of Physics, California Institute of Technology, Pasadena, California 91125, USA}
\author{J. Laurat}
\affiliation{Laboratoire Kastler Brossel, Sorbonne Universit\'e, CNRS, ENS-Universit\'e PSL, Coll\`ege de France, 4 Place Jussieu, 75005 Paris, France}
\author{X. Luan}
\affiliation{Norman Bridge Laboratory of Physics, California Institute of Technology, Pasadena, California 91125, USA}
\author{A. P. Burgers}
\altaffiliation[Present address: Department of Electrical Engineering, Princeton University, Princeton, New Jersey 08540, USA]{}
\affiliation{Norman Bridge Laboratory of Physics, California Institute of Technology, Pasadena, California 91125, USA}
\author{Z. Qin}
\affiliation{Norman Bridge Laboratory of Physics, California Institute of Technology, Pasadena, California 91125, USA}
\affiliation{State Key Laboratory of Quantum Optics and Quantum Optics Devices, Institute of Opto-Electronics, Shanxi University, Taiyuan 030006, China}
\author{H. J. Kimble}
\email{hjkimble@caltech.edu}
\affiliation{Norman Bridge Laboratory of Physics, California Institute of Technology, Pasadena, California 91125, USA}
%


\date{\today}

\pacs{Valid PACS appear here}
%




\begin{abstract}
Spatially structured light has opened a wide range of opportunities for enhanced imaging as well as optical manipulation and particle confinement. Here, we show that phase-coherent illumination with superpositions of radial Laguerre-Gauss (LG) beams provides improved localization for bright optical tweezer traps, with narrowed radial and axial intensity distributions. Further, the Gouy phase shifts for sums of tightly focused radial LG fields can be exploited for novel phase-contrast strategies at the wavelength scale. One example developed here is the suppression of interference fringes from reflection near nano-dielectric surfaces, with the promise of improved cold-atom delivery and manipulation.
\end{abstract}

\maketitle

Structuring of light has provided advanced capabilities in a variety of research fields and technologies, ranging from microscopy to particle manipulation \cite{Allen2003,Ashkin2006,Grier2003,Zhan2009}. Coherent control of the amplitude, phase, and polarization degrees of freedom for light enables the creation of engineered intensity patterns and tailored optical forces. In this context, Laguerre-Gauss (LG) beams have been extensively studied. Among other realizations, tight focusing with subwavelength features was obtained with radially polarized beams \cite{Dorn2003,Wang2008}, as well as with opposite orbital angular momentum for copropagating fields  \cite{Wozniak2016}. LG beams have also attracted interest for designing novel optical tweezers \cite{Padgett2011,Franke2017,Babiker2019}. Following the initial demonstration of a LG-based trap for neutral atoms \cite{Kuga1997}, various configurations have been explored, including 3D geometries with ``dark'' internal volumes \cite{Chaloupka97,Ozeri99, Arlt2000, Arnold2012} for atom trapping with blue-detuned light \cite{Xu2010,Barredo2020}.

For these and other applications of structured light, high spatial resolution is of paramount importance. However in most schemes, resolution transverse to the optic axis largely exceeds that along the optic axis. For example, a typical bright optical tweezer formed from a Gaussian beam with wavelength $\lambda=\SI{1}{\micro\meter}$ focused in vacuum to waist $w_0=\SI{1}{\micro\meter} $ has transverse confinement $w_0$ roughly $3\times$ smaller than its longitudinal confinement set by the Rayleigh range $z_R = \pi w_0^2/\lambda$. One way to obtain enhanced axial resolution is known as 4$\pi$ microscopy \cite{Hell92,Bokor2004}, for which counterpropagating beams form a standing wave with axial spatial scale of $\lambda/2$ over the range of $z_R$. However, 4$\pi$ microscopy requires interferometric stability and delicate mode matching. Another method relies on copropagating beams each with distinct Gouy phases \cite{Boyd80,Steuernagel2005,Birr2017}, which was proposed and realized mostly in the context of dark (i.e., blue-detuned) optical traps, either with two Gaussian beams of different waists or offset foci \cite{Isenhower2009,Teich2003}, or with LG modes of different orders \cite{Arlt2000,Freegarde2002}. However, for bright (i.e, red-detuned) trap configurations, a comparable strategy has remained elusive. 

\begin{figure}[!t]
\centering
\includegraphics[width=\figwidth]{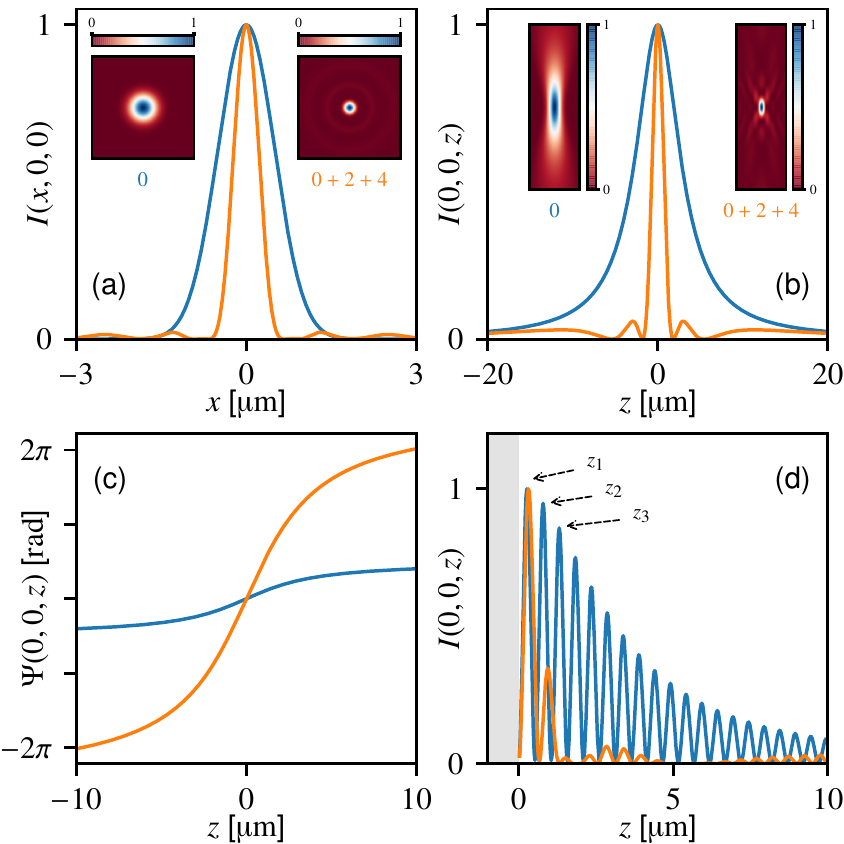}
\vspace{-4mm}
\caption{Comparison between the fundamental Gaussian mode \mbox{$\vec{E}_{0}$} (blue) and the superposition of radial $p$ modes \mbox{$\vec{E}_{\Sigma}$} (orange) with $p={0,2,4}$. The plots are calculated for the paraxial case, with $w_0=\SI{1}{\micro\meter}$ and $\lambda = \SI{1}{\micro\meter}$. (a) $x$-line cut transverse intensity profiles. Insets provide the x-y distribution in the focal plane. (b) $z$-line cut axial intensity profiles. Insets correspond to the x-z distribution in the $y=0$ plane. (c) Gouy phases $\Psi_{0}$ (blue) for \mbox{$\vec{E}_{0}$} and $\Psi_{\Sigma}$ (orange) for \mbox{$\vec{E}_{\Sigma}$} along the optical axis $z$. (d) Reflection fringes due to a semi-infinite planar surface (grey), with amplitude reflection coefficient $r=-0.8$ and focus at the surface $z=0$. $z_i$ indicate successive maxima. All intensities are normalized to their peak values in (a,b,d).}
\label{fig:1}
\vspace{-4mm}
\end{figure}

\begin{figure*}[!t]
\centering
\includegraphics[width=\figwidth]{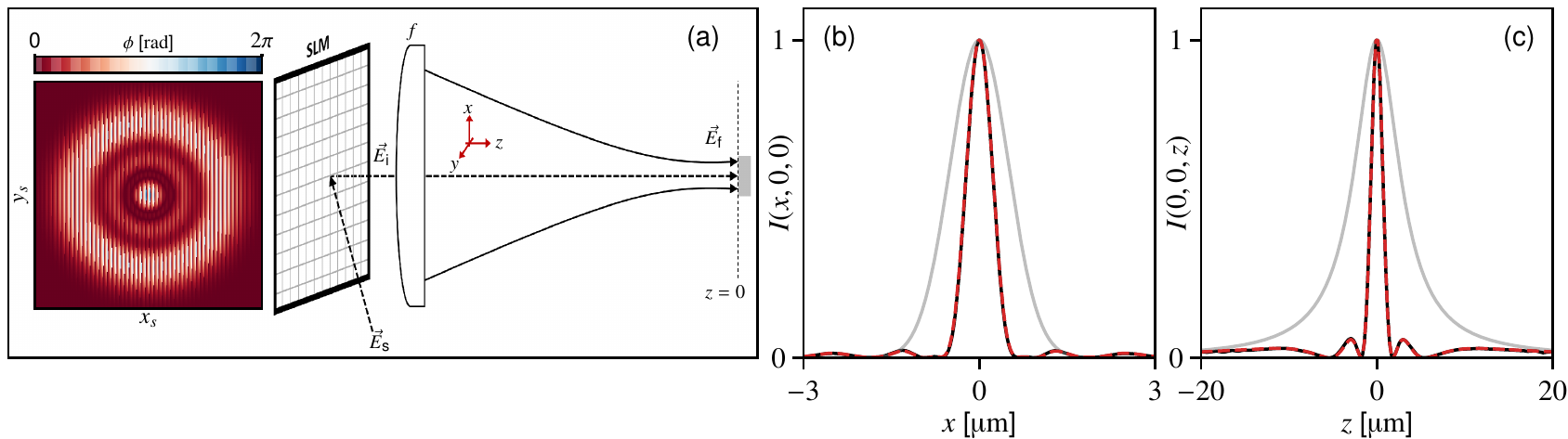}
\vspace{-4mm}
\caption{(a) \textit{left-} Calculated transverse phase profile $\phi$ applied to the spatial light modulator (SLM) to generate the field \mbox{$\vec{E}_{\Sigma}$}. \textit{right-} An incident Gaussian source field $\vec{E}_\text{s}$ is incident on the SLM. The first order diffracted field $\vec{E}_\text{i}$ on the exit plane of the SLM is then focused by an objective lens with effective focal length $f$ to form the field $\vec{E}_\text{f}$ in the focal plane at $z=0$. (b) Line cuts along $x$ of $|\vec{E}_\text{f}|^2$ in the focal plane for modulation of the SLM with $\phi(x_s,y_s)$ calculated to generate \mbox{$\vec{E}_{\Sigma}$} (red solid line), ideal target intensity \mbox{$|\vec{E}_{\Sigma}|^2$} (black dashed line), and Gaussian intensity \mbox{$|\vec{E}_{0}|^2$} (gray line). (c) As in (b), but for line cuts along $z$ with $x=y=0$. }
\label{fig:2}
\vspace{-4mm}
\end{figure*}

In this article, we show that superpositions of purely radial LG modes can lead to reduced volume for bright optical traps. We also provide a scheme for implementation by way of a spatial light modulator (SLM) for beam shaping extended beyond the paraxial approximation into a regime of wavelength-scale traps. Significantly, apart from reduced trap volume, our study highlights differential Gouy phase shifts at the wavelength scale as a novel tool for imaging. An application is the strong suppression of interference fringes from reflections of optical tweezers near surfaces of nanophotonic structures, thereby providing a tool to integrate cold-atom transport and nanoscale quantum optics, a timely topic of paramount importance for the development of the waveguide QED research field \cite{Kimble:18}.

\vspace{-2mm}
\section{Laguerre-Gauss superpositions in the paraxial limit}
\vspace{-2mm}

To gain an intuitive understanding, we first consider superpositions of LG modes within the familiar paraxial approximation. The positive frequency components of the electric field are denoted by \mbox{$\vec{E}_{p,i} =  \vec{x} u_{p}(x,y,z;w_i) e^{-ikz} $} with $x$-oriented linear polarization and propagation directed towards negative $z$ values with longitudinal wave-vector $k>0$. The cylindrically-symmetric complex scalar amplitude $u_{p}$ for LG beams is as in \cite{Allen1992,bookSiegman}, and given explicitly in \cite{SM}. The parameter $w_i$ denotes the waist, i.e., $1/e^2$ intensity radius at $z=0$ for a $p=0$ Gaussian beam. The azimuthal mode number $l$ is dropped with $l=0$ throughout (i.e., pure radial LG beams with radial number $p$). For a given optical frequency, the phase of the field relative to that of a plane wave propagating along $-z$ (i.e., the Gouy phase \cite{Boyd80,Steuernagel2005,Birr2017}) is given by $\Psi_p (z) = \textrm{arg}( \vec{E}_{p,i}\cdot \vec{x} e^{ikz}) = (2p+1) \arctan(z/z_{R,i})$, with the Rayleigh range $z_{R,i} = \pi w_i^2/\lambda$.

Although we have analyzed diverse superpositions of radial LG modes, for clarity we confine our discussion here to the particular superposition $\vec{E}_{\Sigma}=\vec{E}_{0}+\vec{E}_{2}+\vec{E}_{4}$ due to its improvement in atom delivery. For example, the coherent superposition $\vec{E}_{0}+\vec{E}_{6}$ gives a narrower axial focal width as compared to that of $\vec{E}_{0}+\vec{E}_{2}+\vec{E}_{4}$. However, this comes at the price of strong axial sidelobes which is a hindrance for the presented atom delivery scheme due to significant revivals of reflection fringing. For the sole purpose of free-space trapping, note that the phase modulation strategy illustrated in this work is deterministic. Cold atoms could be first loaded in a conventional single p=0 tweezer (absent of intensity side-lodes) and then progressively turning on other p-mode components.

Fig.~\ref{fig:1}(a, b) provide the calculated intensity distributions for the fundamental Gaussian mode \mbox{$\vec{E}_{0}$} (blue) and for the superposition  (orange), along the $x$-axis in the focal plane and along the $z$-propagation axis, respectively. As shown by the line cuts and insets in Fig.~\ref{fig:1} (a, b), there is a large reduction in focal volume $V_{\Sigma}$ for \mbox{$|\vec{E}_{\Sigma}|^2$} relative to $V_{0}$ for \mbox{$|\vec{E}_{0}|^2$}. Here, $V = \Delta x \Delta y \Delta z$, with $\Delta x,\Delta y, \Delta z$ taken to be the full widths at half maxima for the intensity distributions along $x,y,z$ in Fig.~\ref{fig:1}(a, b), leading to $V_{0}/V_{\Sigma} \simeq 22$ where $V_{0} = \SI{8.6}{\cubic\micro\meter}$ and $V_{\Sigma} = \SI{0.39}{\cubic\micro\meter}$ as detailed in \cite{SM}. 

Recall that the root-mean-square radial size $\sigma_p$ of the beam intensity increases as $\sigma_p = w_i\sqrt{2p+1}$ \cite{Phillips1983}, associated with the LG basis scale parameter $w_i$ (i.e., fixed Rayleigh range). This leads to a larger divergence angle for higher radial number $p$.
Therefore, transverse clipping and the impact of diffraction effects due to the constraints of finite aperture stop sizes in any realistic imaging lens system (e.g., finite lens numerical aperture, pupil radius) needs to be included, and is thus analyzed further below.

Also relevant is that individual $u_p$ modes have identical spatial profiles $|u_p(0,0,z)|$ along $z$. The reduced spatial scale for the superposition \mbox{$\vec{E}_{\Sigma}$} results from the set of phases $\{\Psi_p (z)\}$ for $p=0,2,4$, with Gouy phases for the total fields \mbox{$\vec{E}_{0}$} and \mbox{$\vec{E}_{\Sigma}$} shown in Fig.~\ref{fig:1}(c). The Gouy phase for \mbox{$\vec{E}_{\Sigma}$} also leads to suppressed interference fringes in regions near dielectric boundaries as shown in Fig.~\ref{fig:1}(d).

Beyond volume, a second metric for confinement in an optical tweezer is the set of oscillation frequencies for atoms trapped in the tweezer's optical potential. Trap frequencies for Cs atoms localized within tweezers formed from \mbox{$\vec{E}_{0}$} and \mbox{$\vec{E}_{\Sigma}$} as in Fig. \ref{fig:1}(a,b) are presented in \cite{SM}, with significant increases for \mbox{$\vec{E}_{\Sigma}$} as compared to \mbox{$\vec{E}_{0}$}. The values for trap volume and frequency are provided later with the full model.

\vspace{-2mm}
\section{Field superpositions generated with a Spatial Light Modulator}
\vspace{-2mm}

Various methods have been investigated to produce LG beams with high purity \cite{Ando2009}. A relatively simple technique consists of spatial phase modulation of a readily available Gaussian source beam with a series of concentric circular two-level phase steps to replicate the phase distribution of the targeted field $\vec{E}_{p_{target}}$ with $p_{target}>0$ \cite{Arlt1998}. The maximum purity for this technique is $\sim0.8$, with the deficit of $\sim 0.2$ due to the creation of $p$ components other than the single $p_{target}$. Moreover, it is desirable to generate not only high purity LG beams for a single $p_{target}$ but also arbitrary coherent sums of such modes, as for \mbox{$\vec{E}_{\Sigma}$}. Rather than generate separately each component from the set of required radial modes $\{p\}_{target}$, here we propose a technique with a single SLM that eliminates the need to coherently combine multiple beams for the set $\{p\}_{target}$. Our strategy reproduces simultaneously both the phase and the amplitude spatial distribution of the desired complex electric field (and in principle, the polarization distribution for propagation phenomena beyond the scalar field approximation).

Fig.~\ref{fig:2} shows numerical results for a Gaussian source field $\vec{E}_\text{s}$ input to a SLM to create the field $\vec{E}_\text{i}$ leaving the SLM. $\vec{E}_\text{i}$ is then focused by an ideal thin spherical lens and propagated to the focal plane at $z=0$ by way of the Fresnel-Kirchhoff scalar diffraction integral.

Fig.~\ref{fig:2}(a) illustrates our technique for the case of the target field \mbox{$\vec{E}_{\Sigma}$}. Amplitude information for the sum of complex fields comprising \mbox{$\vec{E}_{\Sigma}$} is encoded in a phase mask by contouring the phase-modulation depth of a superimposed blazed grating as developed in \cite{Davis1999,Bolduc2013}. For atom trapping applications with scalar polarizability, the tweezer trap depth is proportional to the peak optical intensity in the focal plane, where for the coherent field superposition \mbox{$\vec{E}_{\Sigma}$}, the peak intensity reaches a value identical to that for \mbox{$\vec{E}_{0}$} at $1/3$ of the invested trap light power, which helps to mitigate losses associated with the blazed grating. We remark that it is not crucial to convert from \mbox{$\vec{E}_{0}$} with simultaneous amplitude and phase modulation strategies, e.g., consider flat-top beams \cite{Ando2009}.

The resulting intensity distributions in the focal plane are plotted in Fig.~\ref{fig:2}(b,c) (red solid) for comparison with the ideal $\vec{E}_\text{i} =\vec{E}_{0}$ (grey solid) and ideal $\vec{E}_\text{i} = \vec{E}_{\Sigma}$ (black dashed). These results are encouraging for our efforts to experimentally generate tightly focused radial LG superpositions (see \cite{SM} for initial laboratory results).

\begin{figure}[t!]
\vspace{-0mm}
\centering
\includegraphics[width=\figwidth]{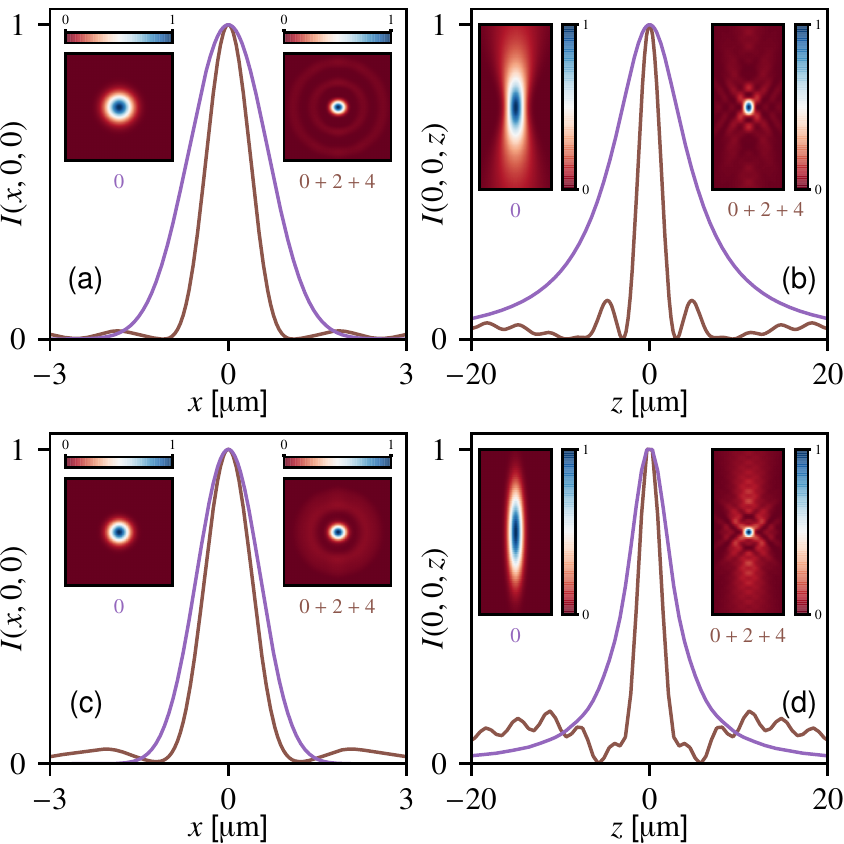}
\vspace{-3mm}
\caption{Focused intensity distributions calculated within the vectorial Debye approximation for $x$-polarized inputs \mbox{$\vec{E}_{0}$} (violet) and \mbox{$\vec{E}_{\Sigma}$} (brown). The numerical aperture is $\textrm{NA}=0.7$ and two filling factor values are compared: For $F_0=0.35$ (a, b). (a) $x$-line cut transverse intensity profiles. The insets provide the x-y intensity distribution in the focal plane $z=0$. (b) $z$-line cut axial intensity profiles. The insets correspond to the x-z distribution. For $F_0=0.45$ (c,d). Plotted intensities for inputs \mbox{$\vec{E}_{0}$} (violet) and \mbox{$\vec{E}_{\Sigma}$} (brown) are normalized to their maximum values.}
\label{fig:3}
\vspace{-4mm}
\end{figure}

\vspace{-2mm}
\section{Vector theory of LG superpositions}
\vspace{-0mm}

Figs. \ref{fig:1} and \ref{fig:2} provide a readily accessible understanding of focused LG mode superpositions within the paraxial approximation. To obtain a more accurate description for tight focusing on a wavelength scale, we next consider a vector theory. Using the vectorial Debye approximation \cite{Richards1959,NovotnyBook} and an input field \mbox{$\vec{E}_{0}$} with waist $w_0 \gg \lambda$ and polarization aligned along the $x$-axis, we calculate the field distribution at the output of an aplanatic objective with fixed numerical aperture $\textrm{NA}={\sin\theta_\text{max}}$.$^{\dag\dag}$

\footnotetext{\small $^{\dag\dag}$ In the paraxial limit with $l=0$, LG modes are completely specified by wavelength, beam waist, and mode order $p$, including beam divergence and rms intensity radius, as applied to the source and input fields. However, for wavelength-scale focusing and finite apertures, the full vector theory is required with more complex parametric dependencies.}

By convention, the ratio of input waist $w_{0,\text{in}}$ to the pupil radius $R_p$ is called the filling factor $F_0 \equiv \frac{w_{0,\text{in}}}{R_p}$, where $R_p = f{\times}\mathrm{NA}$ for focal length $f$. $F_0$ is an important parameter for focusing LG beams at finite aperture, and different filling factors may have very different beam shapes. The curves in Fig.~\ref{fig:3} are calculated numerically using the Debye-Wolf vector theory for filling factors $F_0 =0.35$ and $F_0 =0.45$, each with numerical aperture $\textrm{NA}=0.7$. These parameters provide $1/e^2$ intensity radii $w_{e2}=\{1.3, 1.0\}$ $\mu$m in the focal plane with the input \mbox{$\vec{E}_{0}$} for $F_0 =\{0.35, 0.45\}$, respectively.

For $p=0$ and $F_0=0.35$, the intensity profiles in both radial and axial directions in the focal plane (violet curves in Fig. \ref{fig:3}(a, b)) are quite similar to those in Fig.~\ref{fig:1}(a,b). Likewise, for input of the `0+2+4' superposition at the same filling factor $F_0=0.35$ (brown curves in Fig. \ref{fig:3}(a, b)), the intensity profiles are again similar to Fig.~\ref{fig:1}(a,b) and evidence reductions in both transverse and longitudinal widths relative to the $p=0$ input even in the vector theory with wavelength-scale focusing.

More quantitatively, with $F_0=0.35$ the FWHMs for the `0+2+4' superposition input are $\Delta x_\Sigma = \SI{0.84}{\micro\meter}$,  $\Delta y_\Sigma = \SI{0.72}{\micro\meter}$ and $\Delta z_\Sigma = \SI{2.78}{\micro\meter}$, corresponding to a focal volume $V_{\Sigma} = \SI{1.7}{\cubic\micro\meter}$ for the central peak. For the $p=0$ input with $F_0=0.35$, the FWHMs of the central peaks for each direction are $\Delta x_0 = \SI{1.55}{\micro\meter}$, $\Delta y_0 =  \SI{1.51}{\micro\meter}$ and $\Delta z_0 = \SI{10.3}{\micro\meter}$, corresponding to a focal volume $V_{0} = \SI{24}{\cubic\micro\meter}$. The latter reduces to $V_{0} = \SI{5.62}{\cubic\micro\meter}$ under transverse clipping with $F_0=1$ (see section \ref{Sec:Optimal}).  The factor of focal volumes defined via FWHMs for inputs with $p=0$ and the `0+2+4' superposition is then $V_{0}/V_{\Sigma} \simeq 14$.
Moreover, for red-detuned optical traps associated with the line cuts in Fig. \ref{fig:3}(a, b), we find trap frequencies for input \mbox{$\vec{E}_{\Sigma}$} to be $\omega_x^{\Sigma} =2\pi\times\SI{124}{\kilo\hertz}$ and $\omega_z^{\Sigma} = 2\pi\times\SI{33}{\kilo\hertz}$.

However, increasing of the filling factor beyond $F_0 =0.35$ for the `0+2+4' superposition input does not lead to increases in trap frequencies nor further reductions of the focal volume. As shown by the brown curves in Fig.~\ref{fig:3} (c, d) for filling factor $F_0=0.45$, the central width of the focus is not reduced; rather, the peak of two side lopes increases. This is not the case for the $p=0$ input (violet curves in Fig.~\ref{fig:3} (c, d)), for which the fitted waist $w_0\simeq$ \SI{1}{\micro\meter} for $F_0=0.45$ as compared to $w_0\simeq$ \SI{1.3}{\micro\meter} for $F_0=0.35$. The existence of an `optimal' filling factor for superpositions of LG beams is related to the truncation of the highest order (i.e., $p$ value) in the superposition, which is discussed in Ref. \cite{Luan_thesis}. 

\vspace{-2mm}
\subsection{Filling factor dependence for trap frequencies and dimensions}
An important operational issue for bright tweezer trapping of atoms and molecules is confinement near the intensity maxima shown in Fig. \ref{fig:3}. From various metrics, here we choose to quantify localization by way of trap vibrational frequencies near the bottom of the trapping potential (i.e., the central intensity maximum for a red-detuned trap), which are modified by pupil apodization and diffraction effects for focused radial Laguerre-Gauss beams according to their radial mode number $p$ ~\cite{Haddadi2015}.

\begin{figure}[t!]
\vspace{-2mm}
\centering
\includegraphics[width=\linewidth]{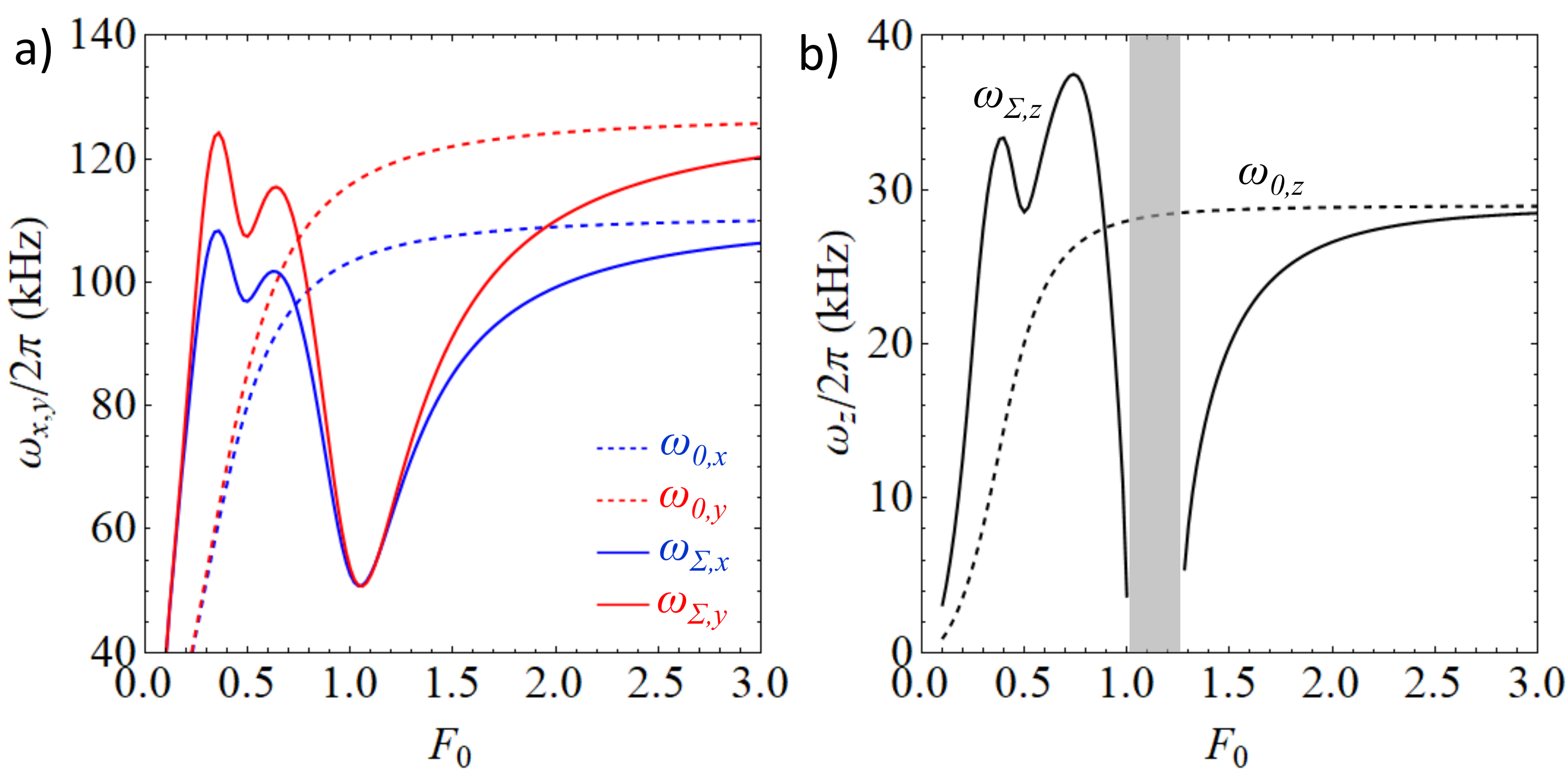}
\vspace{-6mm}
\caption{Radial and axial trap frequencies as a function of filling factor $F_0$ for the $p = 0$ input (dashed) and the ‘0+2+4’ superposition input (solid) (a) trap frequencies in $x$ (blue)  and $y$ (red) directions (b) trap frequencies in $z$ direction (black). The gray shaded area represents regions where the trap center becomes a saddle point with a local intensity minimum. All frequencies are evaluated from the trap origin $(x=y=z=0)$ with normalized trap depth at $U/k_{\text{B}}=1$ mK.}
\label{fig:4}
\vspace{-4mm}
\end{figure}

As an example, we show in Fig. \ref{fig:4} the variation of trap frequency near the trap minimum (intensity maximum) around $x,y,z=0$ for the $x$-polarized input field distributions \mbox{$\vec{E}_{0}$} and \mbox{$\vec{E}_{\Sigma}$} as a function of the objective lens filling factor $F_0$ within the vectorial Debye propagation model~\cite{Richards1959,NovotnyBook}. The angular trap frequencies $\omega_x$, $\omega_y$ and $\omega_z$ are obtained by fitting the trap minimum at $z=0$ to a harmonic potential and extracting the frequency. Fig. \ref{fig:4}(a) corresponds to the transverse trap frequencies $\omega_x$ and $\omega_y$ while Fig. \ref{fig:4}(b) is for the axial frequency $\omega_z$ with corresponding intensity distributions for $F_0=0.35,$ $ 0.45$ shown in Fig. \ref{fig:3}. The grey region in Fig. \ref{fig:4}(b) arises when the curvature at $z=0$ becomes anti-trapping for $1.0\lesssim F_0\lesssim1.26$ with the trap minimum located away from the origin. Plots showing the evolution of the trap around these filling factors can be found in \cite{SM}.

Note that a choice around the local extremum $F_0 \sim 0.35$ not only alleviates practical requirements of the objective lens (e.g., focal length and working distance) but also permits focused fields not dominated by diffraction losses. The reduction of both transverse and longitudinal intensity widths for input \mbox{$\vec{E}_{\Sigma}$} relative to \mbox{$\vec{E}_{0}$} displayed in Fig. \ref{fig:3} are now evident in Fig. \ref{fig:4} for trapping frequencies even in the vector theory with wavelength-scale focusing. For well-chosen filling factors $F_0$, $\omega_z$ for \mbox{$\vec{E}_{\Sigma}$} can be larger than any possible value for $\omega_z$ achieved with the \mbox{$\vec{E}_{0}$} beam (no matter the value of $F_0$).

\vspace{-2mm}
\subsection{Polarization ellipticity for tight focusing}

\begin{figure}[t!]
\vspace{-3mm}
\centering
\includegraphics[width=\linewidth]{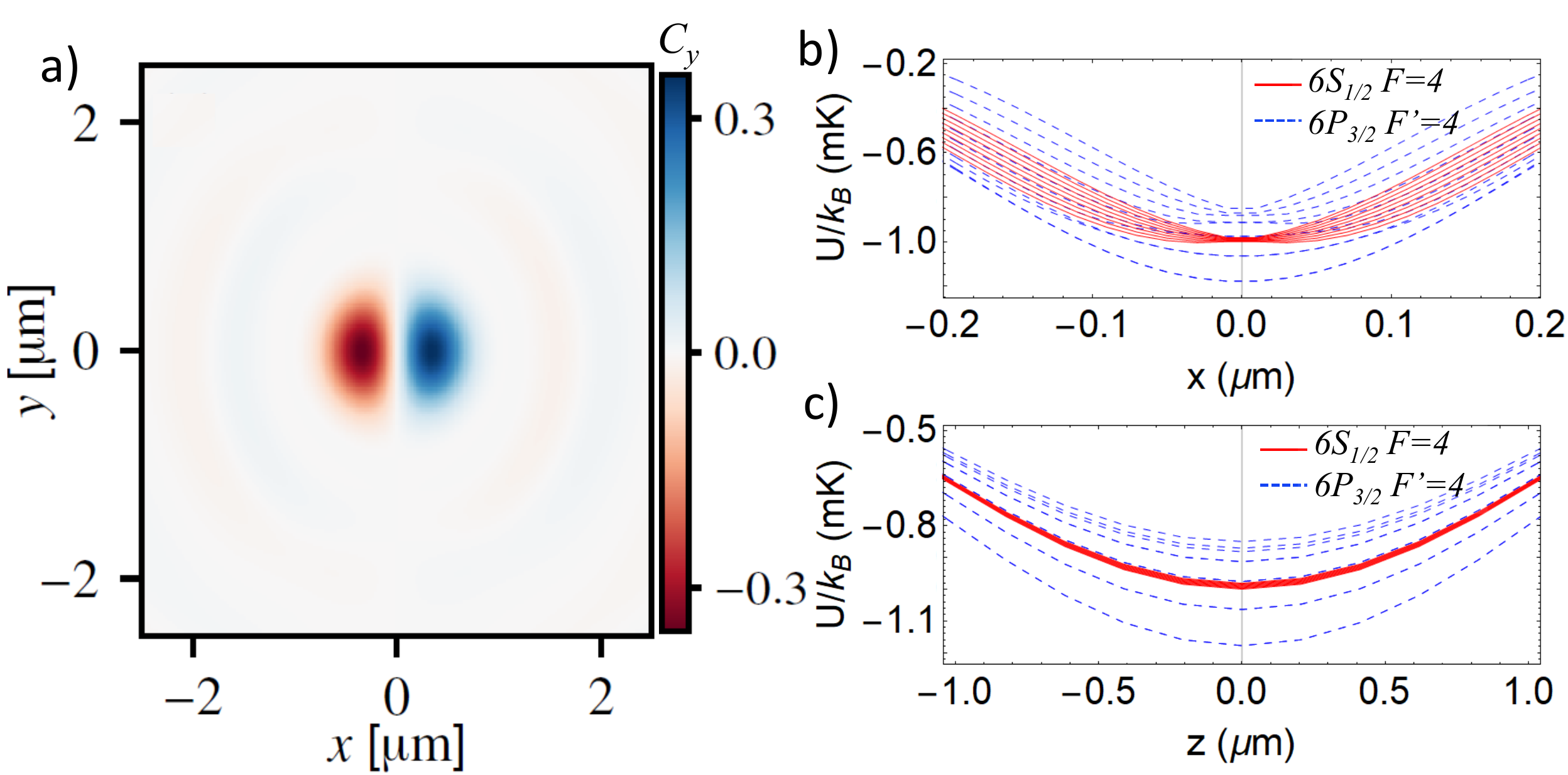}
\vspace{-5mm}
\caption{Polarization ellipticity and vector light shift for the `0+2+4' input. (a) Polarization ellipticity $C_y$ in the focal plane for the `0+2+4' input with $\textrm{NA}=0.7$  and $f_0=0.35$. (b, c) the light shifts for a Cs atom at magic wavelength 935.7 nm with trap depth $U/k_B=$ 1 mK for transverse (b) and axial direction (c). The dashed lines indicate the $m_F$ levels in $6S_{1/2}$, $F=4$ ground state (red dashed) and in $6P_{3/2}$, $F'=4$ excited state (blue dashed). In (b), we can see the ground state trap is shifted away for the center by $\delta x\sim 30$ nm for the $m_F=4$ sublevel.}
\label{fig:5}
\vspace{-4mm}
\end{figure}

Necessarily, tight focusing of optical fields is accompanied by a longitudinal polarization component, which requires a description beyond the atomic scalar polarizability and which results in spatially-dependent elliptical polarization and to dephasing mechanisms for atom trapping \cite{Kuhr2005,ThompsonPRL:13,Goban2012Demonstration,hummer19}.
Given the local polarization vector $\hat{\epsilon}$, one can define the vector $\mathbf{C}=\textrm{Im}(\hat{\epsilon}\times\hat{\epsilon^{\star}})$, which measures the direction and degree of ellipticity.  $|\mathbf{C}| = 0$ corresponds to linear polarization while $|\mathbf{C}| = 1$ for circular polarization. Fig.~\ref{fig:5} (a) provides $C_y$ in the focal plane for the `0+2+4' superposition input. Due to tighter confinement, the polarization gradient reaches $dC_y/dx=1.6/\mu $m for `0+2+4' superposition input, to be compared to $0.4/\mu $m for the $p=0$ input \mbox{$\vec{E}_{0}$}. 


We can further quantify the impact of this ellipticity for trapping atoms by the light shifts (scalar, vector and tensor shifts) of the `0+2+4' superposition for trapping the Cs atom, as shown in Fig.~\ref{fig:5} (b, c).  Here, we choose the wavelength at a magic wavelength of Cs ($\lambda=935.7$ nm) with a given trap depth $U/k_B=1$ mK (for $\textrm{NA}=0.7$ and $F_0=0.35$). Vector light shifts are clearly observed in the transverse direction in Fig.~\ref{fig:5} (b). The trap centers for different $m_F$ levels in $6S_{1/2}$, $F=4$ ground state are shifted away from $x=0$ by $\delta x\sim 30$ nm. As the vector light shift is equivalent to a magnetic field gradient along x direction, it can be suppressed in experiment by an opposite magnetic gradient as demonstrated in Ref.~\cite{ThompsonPRL:13}.  


\section{Optimal filling factors}\label{Sec:Optimal}

\begin{figure}[htbp]
\centering
\vspace{-2mm}
\includegraphics[width=\linewidth]{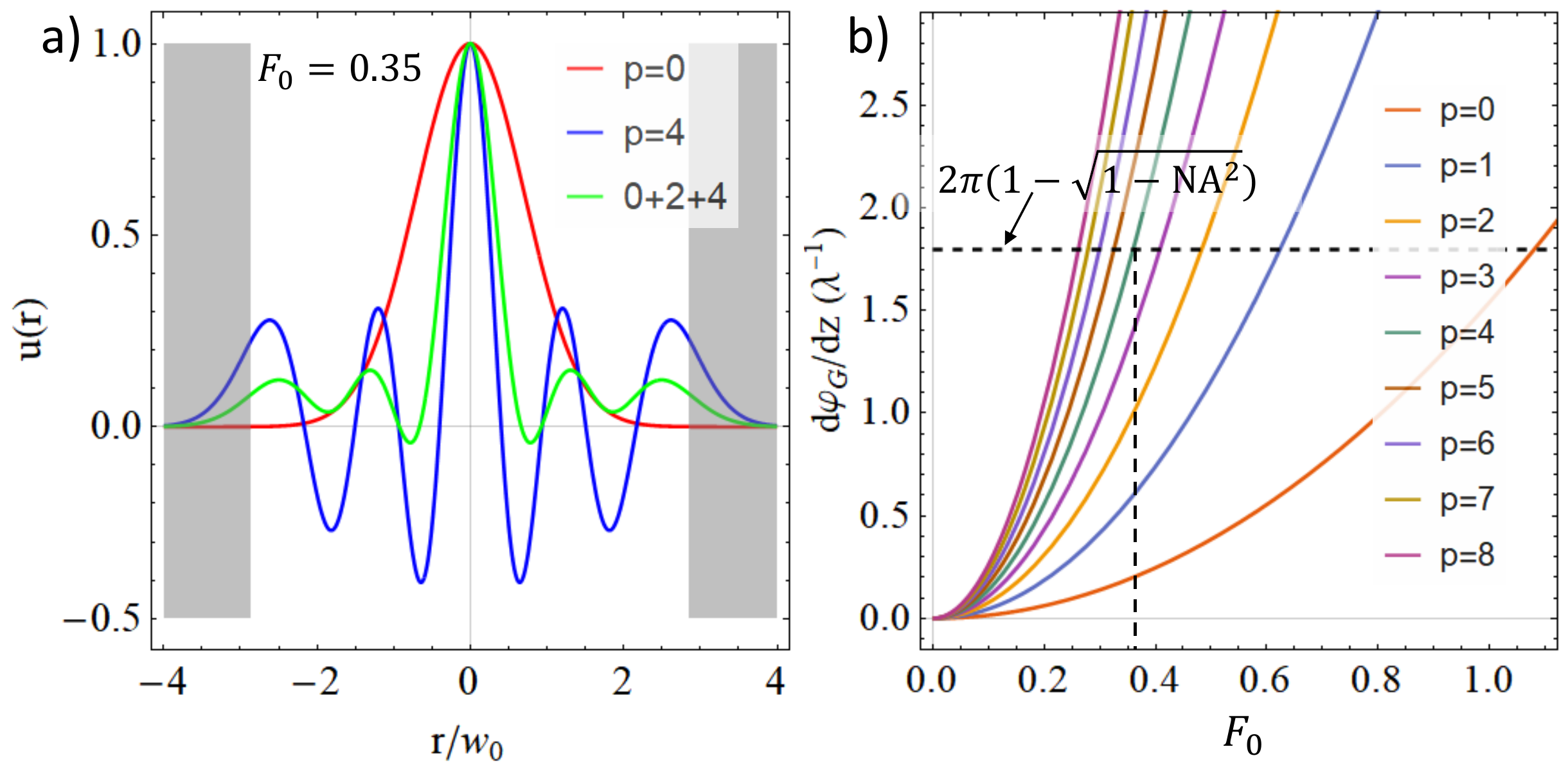}
\caption{Interpretation of optimal filling factor $F_0$ for focusing LG beams with finite aperture objective \cite{Luan_thesis}. (a) Electric field amplitude of $p=0$ (red curve), $p=4$ (blue curve) and `0+2+4' superposition (green curve) at filling factor $F_0=0.35$. The gray shade region represents the physical cutoff from the entrance pupil of the objective with $\textrm{NA}=0.7$. (b) Phase gradient of the focused field for different LG beam inputs assuming $\textrm{NA}=1$. The horizontal dashed line indicates the maximum available phase gradient from a finite objective with $\textrm{NA}=0.7$. The crossing of the horizontal dashed line and phase gradient for $p=4$ LG beam (green curve) corresponds to a filling factor $F_0\simeq 0.35$.}
\label{fig:6}
\end{figure}

As already shown in Fig.~\ref{fig:4} and discussed in the previous section, the truncation of LG beams in finite apertures will lead to optimal filling factors for the superposed LG beam input such as the `0+2+4' superposition. This section focuses on better understanding of this optimization, beginning with Fig.~\ref{fig:6} (a) \cite{Luan_thesis}. Here, we plot the electric field amplitudes for $p=0$, $p=4$ and the `0+2+4' superposition for filling factor $F_0=0.35$. For $F_0=0.35$, the $p=4$ electric field amplitude (blue curve) is already partially truncated by the aperture (gray area). Further increase of the filling factor will misrepresent the $p=4$ LG beam on the input pupil and as a result, the foundation of spatial reduction due to Gouy phase superposition will have to be reconsidered. The pupil apodization effects will modify the spatial properties of the focused radial LG beams according to their radial mode number $p$ \cite{Haddadi2015}. In fact, larger filling factor truncates the LG beams and can generate completely different field profiles (even bottle beams for a single LG $p=1$ mode input). 

\vspace{-2mm}

\vspace{2mm}
Beyond the intuitive picture of truncation of high order LG beams at larger filling factor, we further developed a simple model based on the analysis of Gouy phase to predict the optimal filling factor 
\cite{Luan_thesis}. For focusing a LG beam with waist $w_{0,\text{in}}$ by a lens with focal length $f$ (assuming the input waist is at the lens position), the ABCD matrix from Gaussian optics predicts the input and output waist ($w_0$) are related by
$w_0=f\lambda/\pi w_{0,\text{in}}$. This leads to a Gouy phase as 
\begin{equation}\label{eqn:phase_gradient_NA1}
\begin{aligned}
    \frac{d\psi_G}{dz}&\approx\frac{2p+1}{z_R}\\
    &=\frac{(2p+1)\pi}{\lambda}\frac{w_{0,\text{in}}^2}{f^2}\\
    &=\frac{(2p+1)\pi}{\lambda}F_0^2\text{NA}^2 .
\end{aligned}
\end{equation}
In the last step, we use the fact that $F_0=w_{0,\text{in}}/f\text{NA}$. This suggests for a $\text{NA}=1$ system, the phase gradient increases quadratically with $F_0$ (or input waist $w_{0,\text{in}}$). However, this phase gradient cannot be arbitrarily high for a finite aperture objective. As shown in Ref. \cite{Luan_thesis}, the maximum phase gradient for an objective with fixed $\text{NA}$ is given as
\begin{equation}
   \left(\frac{d\psi_G}{dz}\right)_\text{max}=k(1-\sqrt{1-\text{NA}^2}).
\end{equation}
By setting equal the result from Eq.~\ref{eqn:phase_gradient_NA1} to this maximum phase gradient, we can solve for the optimal filling factor as
\begin{equation}
    F_{0,\text{opt}}=\frac{1}{\text{NA}}\left(\frac{2}{2p+1}\right)^{\frac{1}{2}}\left(1-\sqrt{1-\text{NA}^2}\right)^{\frac{1}{2}}.
\end{equation}
For $\text{NA}=0.7$, $p=4$, this equation predicts an optimal $F_{0,\text{opt}}\simeq0.36$. In Fig.~\ref{fig:6} (b), we show the plot of phase gradient for $p=0$ to $p=8$ based on Eq.~\ref{eqn:phase_gradient_NA1}. The maximum phase gradient for $\text{NA}=0.7$ is also indicated with horizontal dashed lines. The crossing of $\text{NA}=1$ phase gradient (colored curves) with the maximum phase gradient for finite aperture predicts the maximum filling factor for each $p$ mode to preserve its property.  


\vspace{-2mm}
\subsection{Filling factors and trap volumes}
We have investigated more globally parameter sets that could provide `optimal' values for the filling factor $F_0$, where `optimal' would be formulated specific to the particular application, such as imaging or reduced scattering in the focal volume as investigated in the next section. In applying optical tweezers for atom trapping, an `optimal' filling factor might correspond to the highest trap frequency for a given trap depth. It is indeed possible to derive a relation between trap frequency $\omega$ and filling factor $F_0$, at least within the Debye-Wolf formalism, as described in more detail in Ref. \cite{Luan_thesis}.

\begin{figure}[t!]
\vspace{-3mm}
\centering
\includegraphics[width=\linewidth]{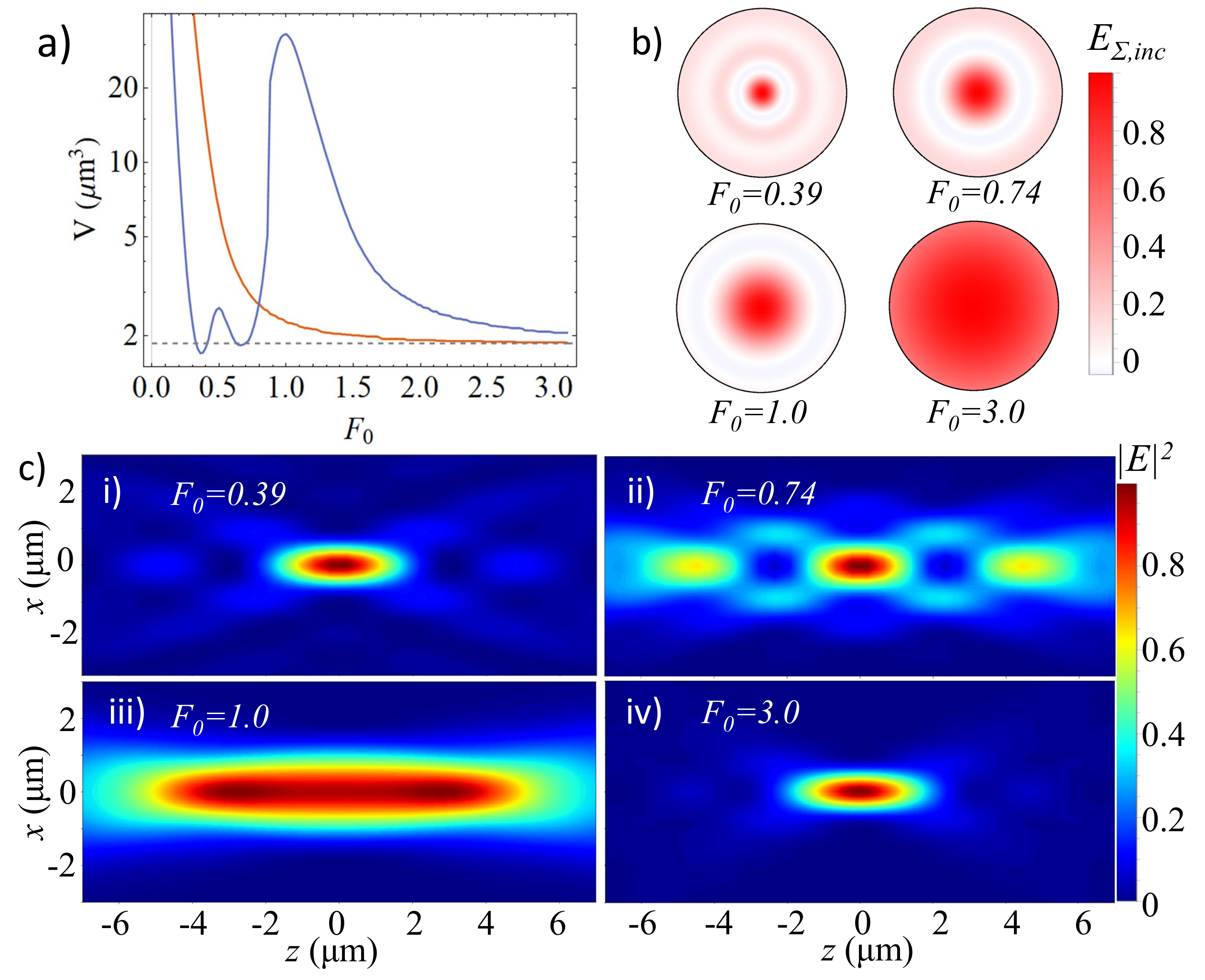}
\caption{Trap volumes and focal-plane intensity profiles for a range of filling factors $F_0$, all with $\text{NA}=0.7$ \cite{Luan_thesis}. (a) Trap volume as a function of filling factor for the $p=0$ input (orange) and the `0+2+4' superposition input (blue). Here the trap volume is defined as $V = \Delta x \Delta y \Delta z$, with $\Delta x,\Delta y, \Delta z$ the full widths at half maxima for the intensity distributions along $x,y,z$. The smallest trap volume we find is $V_{\Sigma} = \SI{1.7}{\cubic\micro\meter}$ at $F_0=0.35$. (b) The input electric field profile for the `0+2+4' superposition at filling factors $F_0=0.39$, $F_0=0.74$, $F_0=1$, and $F_0=3.0$. (c) (i-iv) Intensity profiles near the focus at filling factors $F_ 0=0.39$, $F_0=0.74$, $F_0=1$ and $F_0=3$. Note that the airy rings observed in (i-iii) are still present in (iv) though not revealed due to their small size and the limited contrast.}
\label{fig:7}
\vspace{-4mm}
\end{figure}

Alternatively, for imaging applications, `optimality' might be defined by the value of $F_0$ that achieves the smallest focal volume for a given numerical aperture. Clearly the focal volumes for both imaging and trapping at the wavelength scale are significantly impacted by diffraction and clipping losses of the input field distributions. To investigate this question, Fig. \ref{fig:7}(a) displays volumes $V_0$ and $V_{\Sigma}$ calculated for $x$-polarized inputs of the fields $E_{0}$ and $E_{\Sigma}$, with details of our operational definition of ``volume'' given in \cite{SM}.

Beginning with $F_0 \gg 1$ in Fig. \ref{fig:7}(a), we note that $V_0$ approaches a lower limit that corresponds to the well-known diffraction-limited point-spread-function for a uniformly filled objective of $\text{NA} = 0.7$, which is indeed smaller than $V_{\Sigma}$ for the field $E_{\Sigma}$ in the same limit $F_0 \gg 1$. However, for more modest values $F_0 \simeq 0.3-0.7$, the volume $V_0$ achieved by $E_0$ is significantly larger than $V_{\Sigma}$ for $E_{\Sigma}$ if one matches the input waist at the same objective lens entrance for both fields (diagrams of the input fields at the objective entrance for different filling factors can be found in Fig. \ref{fig:7}(b) and \cite{SM}). Moreover, the volume $V_{\Sigma}$ achieved for $F_0 = 0.38$ is below even the diffraction limit $V_0$ for $F_0 \gg 1$. Importantly, the reduced trap volume for $V_{\Sigma}$ from $E_{\Sigma}$ derives from improved axial localization along $z$ beyond that achievable with $E_0$ for any value of $F_0$ \cite{SM}.

Beyond this general discussion of volume, the behavior of the underlying intensities in the focal volume are complicated for both $E_0$ and $E_{\Sigma}$, with the former well documented in textbooks and research literature and the latter much less so. Hence, in Fig. \ref{fig:7}(b, c) are displayed intensity distributions for $E_{\Sigma}$ (b) across the source aperture and (c) in the focal plane that are much more complicated than those for $E_0$ and which exhibit structure in regions well outside the central maxima (including  strong side lobes and extended axial variations)\cite{SM}. Such side lobes can introduce atom heating for transport of cold atoms in moving tweezers, thereby reducing atom delivery efficiency from free-space to dielectric surfaces. Two other filling factors $F_0=1$ and $F_0=3$ are also presented in Fig.~\ref{fig:7} (b, c). Note that $F_0=1$ in Fig. \ref{fig:7}(c)(iii) corresponds to a flattened trap intensity in the axial direction with properties similar to Bessel beams \cite{Durnin:87:Bessel} with plots of this effect and discussion found in \cite{SM}). Finally, $F_0=3$ in (iv) approaches the limit of a uniform input with a well-known diffraction-limited spot size.

While we have considered here two examples of `optimization' by way of trapping frequencies and volumes in tweezer traps, similar analyses can be formulated to optimize other metrics \cite{Luan_thesis}. Indeed, the systematic search for `optimal' values of $F_0$ briefly described in this section immediately found the peaks shown in Fig. \ref{fig:4} for atom trapping, which we first identified by a considerably more painful random search. Moreover, because trap volumes for focused red traps scale as $V_\text{trap}\propto (\omega_z\omega_x \omega_y )^{-1}$, the expressions for trap frequencies in the axial and transverse directions can be combined to find optimal filling factors $F_0$ to minimize trap volume around the center of a trap for a given input profile $E_\text{in}$ for comparsion to more global measures of volume (e.g., FWHM) \cite{SM}.

\vspace{-2mm}
\section{LG beams reflected from dielectric nanostructures}

Excepting panel (d) in Fig. ~\ref{fig:1}, we have thus far directed attention to free-space optical tweezers for atoms and molecules. However, there are important settings for both particle trapping and imaging in which the focal region is not homogeneous but instead contains significant spatial variations of the dielectric constant over a wide range of length scales from nanometers to microns. Important examples in AMO Physics include recent efforts to trap atoms near nano-photonic structures such as dielectric optical cavities and photonic crystal waveguides (PCWs) \cite{ThompsonPRL:13,Kimble:18,Tiecke:14,Hood:16,Burgers:19,Kim:19}. These efforts have been hampered by large modification of the trapping potential of an optical tweezer in the vicinity of a nano-photonic structure, principally associated with specular reflection that produces high-contrast interference fringes extending well beyond the volume of the tweezer.

The magnitude of the problem is already made clear in the paraxial limit by the blue curve in Panel (d) of Fig. ~\ref{fig:1}. The otherwise smoothly varying tweezer intensities in free-space, shown in Panels (a, b) of Fig. ~\ref{fig:1}, become strongly modulated in Panel (d) by the reflection of the tweezer field from the dielectric surface. Given that the goal for the integration of cold atoms and nanophotonics is to achieve $1\text{D}$ and $2\text{D}$ atomic lattices trapped at distances $z \lesssim \lambda/10$ from surfaces, and that one interference fringe in Fig. \ref{fig:1} spans $\Delta z = \lambda/2$, it is clear that free-space tweezer traps cannot be readily employed for direct transport of atoms along a linear trajectory in $z$ to the near fields of nano-scale dielectrics without implementing more complicated trajectories. These trajectories not only require the tweezer spot to traverse along $z$ but $x$ or $y$ as well \cite{ThompsonSci:2013}. Further insight for direct transport along $z$ is provided by the animations in \cite{SM} for the evolution of the intensity of a conventional optical tweezer as the focal spot is moved from an initial distance  $z_i \gg \lambda$ to a final distance $z_0=0$ at the dielectric surface. Placing atoms at distances $z \lesssim \lambda/10$ from dielectric surfaces is possible by combining LG beam optical tweezers and utilizing guided modes (GM) of the dielectric structure.  These GMs can be configured in such a way to attract the atoms via the dipole force to stable trapping regions z $\lesssim \lambda/10$ from the dielectric. Such trapping configurations are discussed in Ref \cite{Burgers:19}.  

\begin{figure}[t!]
\centering
\includegraphics[width=\linewidth]{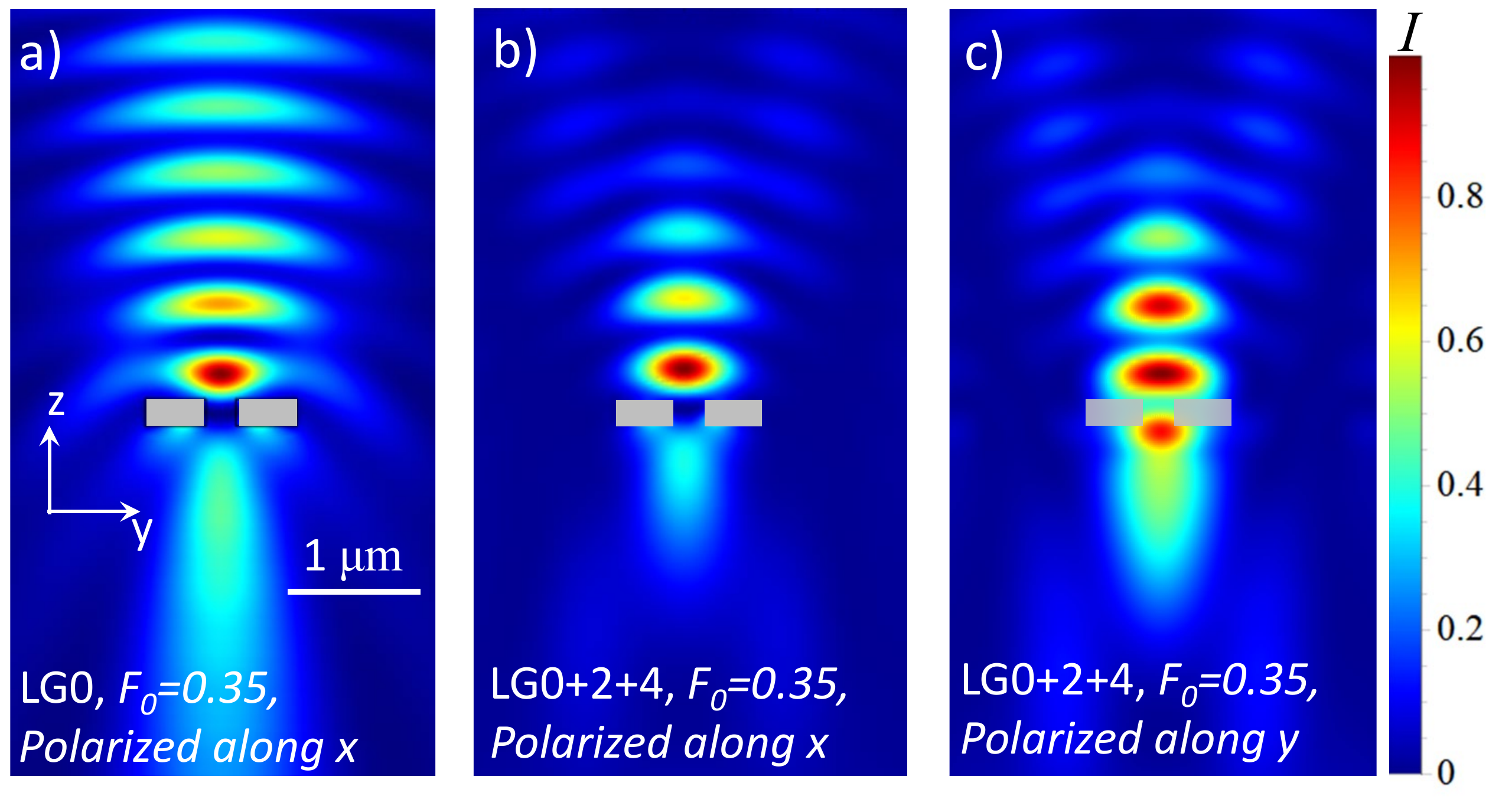}
\caption{Simulation of aligning a tightly focused LG beam to reflect and scatter from an APCW directed out of the page (indicated by the gray rectangle) with focal position aligned to the geometric center of the APCW. (a) For the input field distribution of a $p=0$ Gaussian beam with initial polarization along x, (b) For the input field distribution of the `0+2+4' superposed LG beam with polarization along x, and (c) For the input field distribution of the `0+2+4' superposed LG beam with polarization along y. All three plots are calculated with the background field derived from the Debye-Wolf integral with $\text{NA}=0.7$ and $F_0=0.35$ \cite{Luan_thesis}.}
\label{fig:8}
\vspace{-3mm}
\end{figure}
That said, Fig. \ref{fig:1} d) investigates a strategy to mitigate this difficulty by exploiting the rapid spatial variation of the Gouy phase $\Psi_{\Sigma}$ for the field \mbox{$\vec{E}_{\Sigma}$} as compared to $\Psi_{0}$ for the field \mbox{$\vec{E}_{0}$}. As shown by the orange curve in panel (d), the contrast and spatial extent of near-field interference is greatly reduced for \mbox{$\vec{E}_{\Sigma}$} due to rapid spatial dephasing between input and reflected fields.

To transition this idea into the regime of nanophotonic structures with tightly focused tweezer fields on the wavelength scale, we start with a free-space LG beam in the paraxial limit with waist much larger than the optical wavelength, $w_0 \gg \lambda$. The optical field for this initial LG beam is first `sculpted' with the SLM and then tightly focused as in Fig. \ref{fig:2} with fields in the free-space focal volume calculated from the Debye-Wolf formalism and serving as a background field without scattering. We then solve for the scattered field in the presence of a dielectric nano-structure in the focal volume.

An example to validate directly the possibility of reduced reflection and `fringe' fields for wavelength scale optical tweezers near nanophotonic devices is presented in Fig.~\ref{fig:8} which displays intensity distributions calculated for (a) a focused $p=0$ Gaussian beam input and (b, c) a focused `0+2+4' superposed LG beam aligned to an Alligator Photonic Crystal Waveguide (APCW) \cite{Yu:14} for $\text{NA}=0.7$ and $F_0=0.35$. This result confirms the spatial reduction of ``fringe'' fields from the superposition of LG beams near complex dielectric nanostructures. 

\begin{figure}[t!]
\centering
\includegraphics[width=1.0\linewidth]{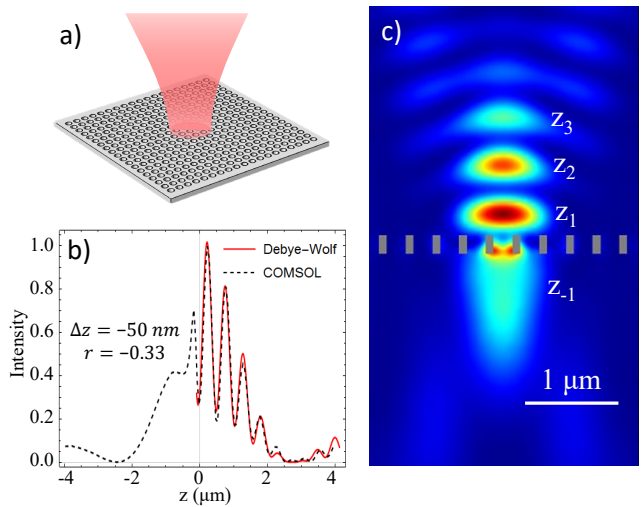}
\caption{(a) Optical tweezer focused on a 2D photonic crystal slab. (b) Fitting of the Comsol simulated field (dashed line) with the Debye-Wolf evaluated field reflecting from a planar surface (red solid line). (c) The Comsol simulated field of the LG `0+2+4' superposition $E_{\Sigma}$ reflected from a $2\text{D}$ PCW comprised of a dielectric slab with a $2\text{D}$ square lattice of holes as in \cite{Yu2019}.}
\label{fig:9}
\end{figure}

We stress that our methods for finding the reflected and scattered fields for nano-photonic devices illuminated by coherent sums of LG fields can be readily extended from $1\text{D}$ to $2\text{D}$ slab PCWs \cite{Yu2019,Tudela2015}. One such result for a $2\text{D}$ square lattice \cite{Yu2019} has been calculated with the vector theory and is displayed in Fig. \ref{fig:9}, again with reduced reflected fields brought by interference from the range of Gouy phases.

\vspace{-2mm}
\section{Atom transport to a photonic crystal}

To investigate the efficiency for atom transport from free-space optical tweezers to reflective traps near dielectric surfaces, we have performed Monte Carlo simulations of atom trajectories by moving a tweezer's focus position from far away ($z=\SI{600}{\micro\meter}$) to the surfaces ($z=\SI{0}{\micro\meter}$) of various nanophotonic devices. These simulations have been carried out for both the paraxial regime and with the full vector theory of Debye-Wolf.

\begin{figure}[t!]
\centering
\includegraphics[width=\figwidth]{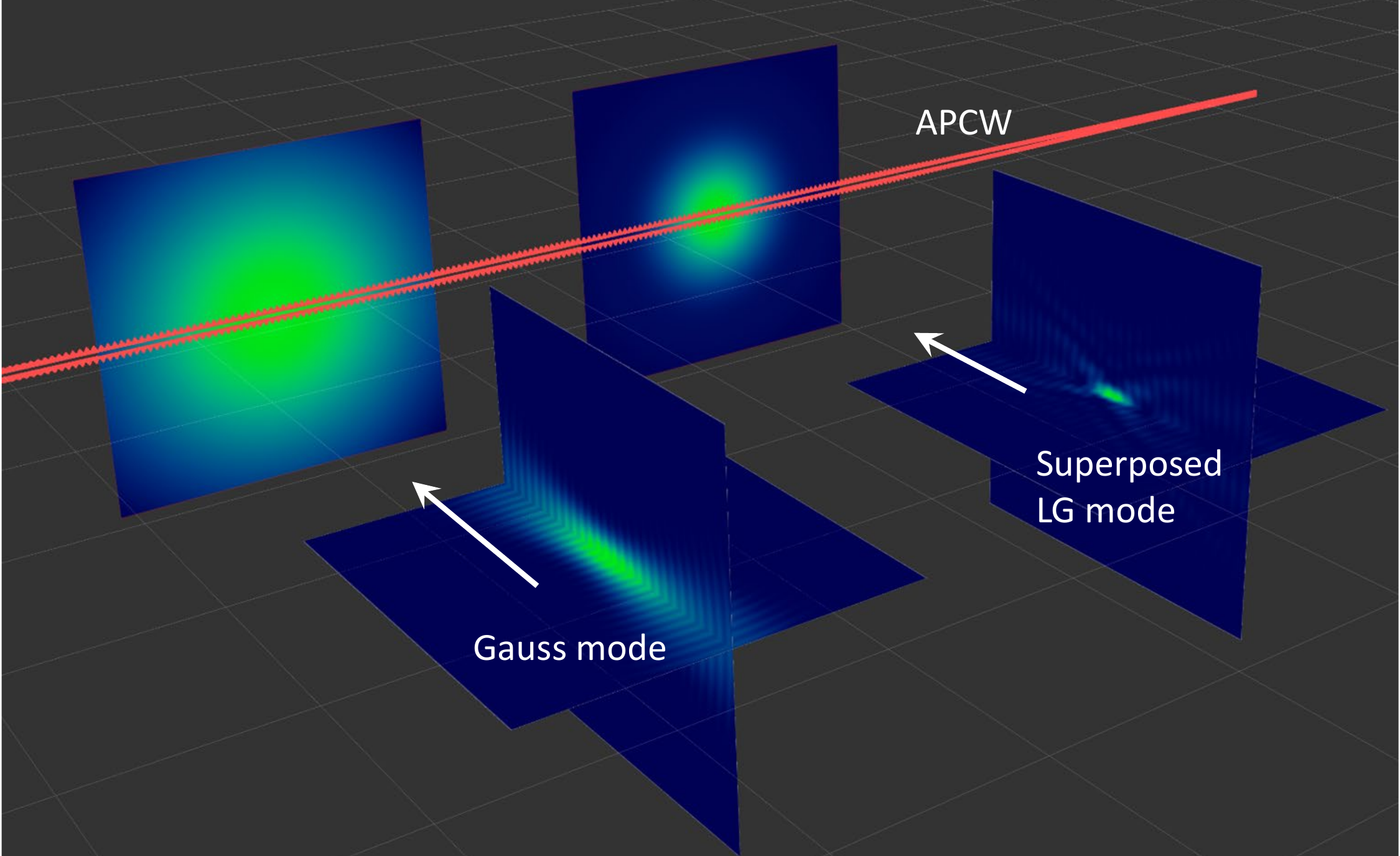}
\vspace{-1mm}
\caption{Single frame from an animation of atom delivery to the Alligator Photonic Crystal Waveguide (in red) by way of a moving optical tweezer. Left: Gaussian beam $E_0$, and Right: Coherent superposition of Laguerre-Gauss beams $E_{\Sigma}$. In the displayed frame, atoms are absent to better highlight the intensity distributions of the two optical tweezers. The white arrows indicate the direction of motion of the tweezer focus as implemented in Refs. \cite{Beguin2020,Luan2020integration} and the grids are 10 $\mu$m$\times$10 $\mu$m. The full animation depicts non-interacting atoms as white `dots' as might have been initially loaded and cooled into the tweezers far from the APCW (20 atoms for each tweezer). A movie can be found at the following link: \url{http://dx.doi.org/10.22002/D1.1446}.
}\label{fig:10}
\vspace{-1mm}
\end{figure}

Fig. \ref{fig:10} and the accompanying animation provide a global view of the intensity distributions for an optical tweezer initially located far from an APCW with then the tweezer focus moved to the surface of the APCW. Two tweezer fields are shown, first for the field $E_0$ for a conventional $p=0$ Gaussian tweezer and second for the unconventional field $E_{\Sigma}$ for the coherent superposition of $p=0,2,4$ beams.

\begin{figure}[t!]
\centering
\includegraphics[width=\figwidth]{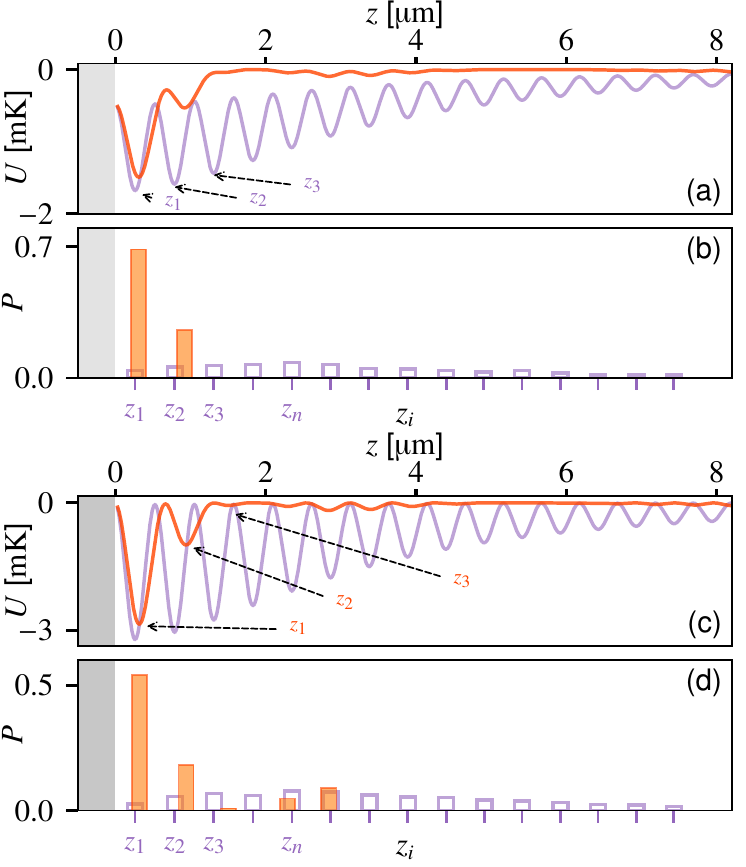}
\vspace{-1mm}
\caption{Results from Monte Carlo simulation of cold atom delivery close to a semi-infinite planar surface with an amplitude reflection coefficient $r=-0.3$ for (a) and (b), and $r=-0.8$ for (c) and (d). (a,c) Optical potentials $U(0,0,z)$ for optical tweezers formed from the fields \mbox{$\vec{E}_{0}$} (violet) and \mbox{$\vec{E}_{\Sigma}$} (red), respectively, as in Fig. 1(d) for focus at $z=0$. (b,d) The final probabilities $P(z_i)$ for delivery of atoms to optical traps centered at positions $z_i$. Atoms are initially loaded into an optical tweezer of depth $U_0=\SI{1}{\milli\kelvin}$ at focal distance $z_{initial}=\SI{600}{\micro\meter}$ from the surface and initial temperature of $\SI{100}{\micro\kelvin}$. The focal plane of the optical tweezer is then scanned from $z_{initial}$ to $z_{final}=\SI{0}{\micro\meter}$. $z_n$ indicates the optical trap formed from $\vec{E}_0$ with highest delivery probability. Full numerical simulations of atom transport can be found in \cite{SM}, and provide the basis for this figure.}
\label{fig:11}
\vspace{-1mm}
\end{figure}

With the overall view in mind from the animation accompanying Fig. \ref{fig:10}, we finally address the question of quantitatively assessing the efficiency of atom transport from free-space tweezers to the near fields of nano-photonic structures for optical tweezers with $E_{\Sigma}$ compared to tweezers with $E_0$. Fig. \ref{fig:11} provides such an assessment of the probabilities for single atoms to be delivered and trapped in surface traps of a reflecting dielectric. The specific choice of reflection coefficient $r=-0.8$ is based on numerical simulations of wavelength-scale tweezer reflection from the nanoscale surface of an Alligator Photonic Crystal Waveguide (APCW) for polarization parallel \cite{Hood:16} to the long axis of the APCW.

Fig. ~\ref{fig:11}(b) confirms that the trap formed by the superposition \mbox{$\vec{E}_{\Sigma}$} (orange histogram) leads to large enhancement in delivery efficiency into near surface traps ($z_1$, $z_2$...) as compared to the very small probability of delivery for the conventional trap formed by \mbox{$\vec{E}_{0}$} (violet histogram). The probability of delivering an atom into the $z_1$ trap with \mbox{$\vec{E}_{\Sigma}$} is $P_{\Sigma}(z_1) \simeq 0.55$ as compared to $P_{0}(z_1) \simeq 0.03$ with \mbox{$\vec{E}_{0}$}. Fig. \ref{fig:11} is from a one-dimensional model of atom transport (i.e., in the optical potential $U(0,0,z)$), and hence provides only a qualitative guide. We have also carried out full 3D simulations for the situation of Fig. \ref{fig:1}(d), with comparable results (e.g., $P_{\Sigma}(z_1) \simeq 0.45$) presented in \cite{SM}.

Moreover, beyond the protocols considered in Fig. \ref{fig:11} and in \cite{SM}, we have found improvements in delivery efficiency by including atom cooling in the simulations at various stages of the transport, as well as applying blue-detuned guided-mode (GM) beams as atoms arrive near the surface to overcome loss due to surface forces such as the Casimir-Polder potential. Finally, to document the robustness of our scheme, results from simulations analogous to those in Fig.~\ref{fig:11} are presented in \cite{SM} for $r=-0.3$ corresponding to an optical tweezer with polarization perpendicular to the long axis of an APCW, for which $P_{\Sigma}(z_1) \simeq 0.68$. 

\vspace{-2mm}
\section{Conclusion and Outlook}
We have proposed coherent superpositions of radial  Laguerre-Gaussian (LG) beams for bright optical tweezers. By way of a vector theory that encompasses diffraction and tight focusing on the wavelength scale, we have investigated new possibilities for reduced trap volumes and increased trapping frequencies for free-space tweezer traps constrained by fixed numerical aperture.  A specific application has been numerically analyzed for the efficient transport of atoms via red-detuned optical tweezers directly to trap sites near the surfaces of nanoscopic dielectric structures. The key feature of our approach is the suppression of interference fringes from reflection near nanoscopic dielectric surfaces. Our goal is to enable a leap forward for cold-atom delivery and manipulation to allow the assembly of 1D and 2D nanoscopic atomic lattices near photonic crystal waveguides by way of the novel optical tweezers that we describe. While bits and pieces of our protocols have appeared in prior papers, to our knowledge, the key aspects of the results presented in our manuscript have not been known heretofore.

Beyond atom trapping with optical tweezers in AMO Physics, we are currently exploring novel imaging techniques with large phase gradients and sub-wavelength scale resolution. For example, Fig. \ref{fig:10} and the accompanying animation document the potential for significantly reduced depth of field (and hence possible improved axial resolution) for illumination and detection by way of coherent superpositions of LG beams, specifically the field $E_{\Sigma}$ relative to the conventional field $E_{0}$.

More generally, the possibility to engineer Gouy phase shifts for sums of tightly focused radial LG fields might extend the range of imaging methods to permit novel phase-contrast microscopy strategies on a sub-wavelength scale, which is an application that we are currently exploring. Beyond engineered nanophotonic structures, the suppression or enhancement of interference from diffuse reflection and scattering in spatially heterogeneous sample volumes (e.g., living cells) is another application under consideration.

\section*{Acknowledgments}
\vspace{-4mm}
The authors thank Robert Boyd and Nick Black for discussions related to SLMs. HJK acknowledges funding from ONR MURI Quantum Opto-Mechanics with Atoms and Nanostructured Diamond Grant No. N000141512761; ONR Grant No. N000141612399; AFOSR MURI Photonic Quantum Matter Grant No. FA95501610323; NSF Grant No. PHY1205729; Caltech KNI. JL acknowledges funding from French-US Fulbright Commission; French National Research Agency (NanoStrong project ANR-18-CE47-0008); R\'egion Ile-de-France (DIM SIRTEQ).

\vspace{-3mm}

\bibliography{mainv3,bibtex_LG}

\begin{thebibliography}{53}%
\makeatletter
\providecommand \@ifxundefined [1]{%
 \@ifx{#1\undefined}
}%
\providecommand \@ifnum [1]{%
 \ifnum #1\expandafter \@firstoftwo
 \else \expandafter \@secondoftwo
 \fi
}%
\providecommand \@ifx [1]{%
 \ifx #1\expandafter \@firstoftwo
 \else \expandafter \@secondoftwo
 \fi
}%
\providecommand \natexlab [1]{#1}%
\providecommand \enquote  [1]{``#1''}%
\providecommand \bibnamefont  [1]{#1}%
\providecommand \bibfnamefont [1]{#1}%
\providecommand \citenamefont [1]{#1}%
\providecommand \href@noop [0]{\@secondoftwo}%
\providecommand \href [0]{\begingroup \@sanitize@url \@href}%
\providecommand \@href[1]{\@@startlink{#1}\@@href}%
\providecommand \@@href[1]{\endgroup#1\@@endlink}%
\providecommand \@sanitize@url [0]{\catcode `\\12\catcode `\$12\catcode
  `\&12\catcode `\#12\catcode `\^12\catcode `\_12\catcode `\%12\relax}%
\providecommand \@@startlink[1]{}%
\providecommand \@@endlink[0]{}%
\providecommand \url  [0]{\begingroup\@sanitize@url \@url }%
\providecommand \@url [1]{\endgroup\@href {#1}{\urlprefix }}%
\providecommand \urlprefix  [0]{URL }%
\providecommand \Eprint [0]{\href }%
\providecommand \doibase [0]{http://dx.doi.org/}%
\providecommand \selectlanguage [0]{\@gobble}%
\providecommand \bibinfo  [0]{\@secondoftwo}%
\providecommand \bibfield  [0]{\@secondoftwo}%
\providecommand \translation [1]{[#1]}%
\providecommand \BibitemOpen [0]{}%
\providecommand \bibitemStop [0]{}%
\providecommand \bibitemNoStop [0]{.\EOS\space}%
\providecommand \EOS [0]{\spacefactor3000\relax}%
\providecommand \BibitemShut  [1]{\csname bibitem#1\endcsname}%
\let\auto@bib@innerbib\@empty
\bibitem [{\citenamefont {Allen}\ \emph {et~al.}(2003)\citenamefont {Allen},
  \citenamefont {Barnett},\ and\ \citenamefont {Padgett}}]{Allen2003}%
  \BibitemOpen
  \bibfield  {author} {\bibinfo {author} {\bibfnamefont {L.}~\bibnamefont
  {Allen}}, \bibinfo {author} {\bibfnamefont {S.~M.}\ \bibnamefont {Barnett}},
  \ and\ \bibinfo {author} {\bibfnamefont {M.~J.}\ \bibnamefont {Padgett}},\
  }\href@noop {} {\emph {\bibinfo {title} {Optical Angular Momentum}}}\
  (\bibinfo  {publisher} {IOP Publishing},\ \bibinfo {year} {2003})\BibitemShut
  {NoStop}%
\bibitem [{\citenamefont {Ashkin}(2006)}]{Ashkin2006}%
  \BibitemOpen
  \bibfield  {author} {\bibinfo {author} {\bibfnamefont {A.}~\bibnamefont
  {Ashkin}},\ }\href@noop {} {\emph {\bibinfo {title} {Optical Trapping and
  Manipulation of Neutral Particles Using Lasers}}}\ (\bibinfo  {publisher}
  {World Scientific},\ \bibinfo {year} {2006})\BibitemShut {NoStop}%
\bibitem [{\citenamefont {Grier}(2003)}]{Grier2003}%
  \BibitemOpen
  \bibfield  {author} {\bibinfo {author} {\bibfnamefont {D.~G.}\ \bibnamefont
  {Grier}},\ }\href {\doibase 10.1038/nature01935} {\bibfield  {journal}
  {\bibinfo  {journal} {Nature}\ }\textbf {\bibinfo {volume} {424}},\ \bibinfo
  {pages} {810} (\bibinfo {year} {2003})}\BibitemShut {NoStop}%
\bibitem [{\citenamefont {Zhan}(2009)}]{Zhan2009}%
  \BibitemOpen
  \bibfield  {author} {\bibinfo {author} {\bibfnamefont {Q.}~\bibnamefont
  {Zhan}},\ }\href {\doibase 10.1364/aop.1.000001} {\bibfield  {journal}
  {\bibinfo  {journal} {Advances in Optics and Photonics}\ }\textbf {\bibinfo
  {volume} {1}},\ \bibinfo {pages} {1} (\bibinfo {year} {2009})}\BibitemShut
  {NoStop}%
\bibitem [{\citenamefont {Dorn}\ \emph {et~al.}(2003)\citenamefont {Dorn},
  \citenamefont {Quabis},\ and\ \citenamefont {Leuchs}}]{Dorn2003}%
  \BibitemOpen
  \bibfield  {author} {\bibinfo {author} {\bibfnamefont {R.}~\bibnamefont
  {Dorn}}, \bibinfo {author} {\bibfnamefont {S.}~\bibnamefont {Quabis}}, \ and\
  \bibinfo {author} {\bibfnamefont {G.}~\bibnamefont {Leuchs}},\ }\href
  {\doibase 10.1103/physrevlett.91.233901} {\bibfield  {journal} {\bibinfo
  {journal} {Physical Review Letters}\ }\textbf {\bibinfo {volume} {91}},\
  \bibinfo {pages} {233901} (\bibinfo {year} {2003})}\BibitemShut {NoStop}%
\bibitem [{\citenamefont {Wang}\ \emph {et~al.}(2008)\citenamefont {Wang},
  \citenamefont {Shi}, \citenamefont {Lukyanchuk}, \citenamefont {Sheppard},\
  and\ \citenamefont {Chong}}]{Wang2008}%
  \BibitemOpen
  \bibfield  {author} {\bibinfo {author} {\bibfnamefont {H.}~\bibnamefont
  {Wang}}, \bibinfo {author} {\bibfnamefont {L.}~\bibnamefont {Shi}}, \bibinfo
  {author} {\bibfnamefont {B.}~\bibnamefont {Lukyanchuk}}, \bibinfo {author}
  {\bibfnamefont {C.}~\bibnamefont {Sheppard}}, \ and\ \bibinfo {author}
  {\bibfnamefont {C.~T.}\ \bibnamefont {Chong}},\ }\href {\doibase
  10.1038/nphoton.2008.127} {\bibfield  {journal} {\bibinfo  {journal} {Nature
  Photonics}\ }\textbf {\bibinfo {volume} {2}},\ \bibinfo {pages} {501}
  (\bibinfo {year} {2008})}\BibitemShut {NoStop}%
\bibitem [{\citenamefont {Wo\'niak}\ \emph {et~al.}(2016)\citenamefont
  {Wo\'niak}, \citenamefont {Banzer}, \citenamefont {Bouchard}, \citenamefont
  {Karimi}, \citenamefont {Leuchs},\ and\ \citenamefont {Boyd}}]{Wozniak2016}%
  \BibitemOpen
  \bibfield  {author} {\bibinfo {author} {\bibfnamefont {P.}~\bibnamefont
  {Wo\'niak}}, \bibinfo {author} {\bibfnamefont {P.}~\bibnamefont {Banzer}},
  \bibinfo {author} {\bibfnamefont {F.}~\bibnamefont {Bouchard}}, \bibinfo
  {author} {\bibfnamefont {E.}~\bibnamefont {Karimi}}, \bibinfo {author}
  {\bibfnamefont {G.}~\bibnamefont {Leuchs}}, \ and\ \bibinfo {author}
  {\bibfnamefont {R.~W.}\ \bibnamefont {Boyd}},\ }\href {\doibase
  10.1103/physreva.94.021803} {\bibfield  {journal} {\bibinfo  {journal}
  {Physical Review A}\ }\textbf {\bibinfo {volume} {94}},\ \bibinfo {pages}
  {021803} (\bibinfo {year} {2016})}\BibitemShut {NoStop}%
\bibitem [{\citenamefont {Padgett}\ and\ \citenamefont
  {Bowman}(2011)}]{Padgett2011}%
  \BibitemOpen
  \bibfield  {author} {\bibinfo {author} {\bibfnamefont {M.}~\bibnamefont
  {Padgett}}\ and\ \bibinfo {author} {\bibfnamefont {R.}~\bibnamefont
  {Bowman}},\ }\href {\doibase 10.1038/nphoton.2011.81} {\bibfield  {journal}
  {\bibinfo  {journal} {Nature Photonics}\ }\textbf {\bibinfo {volume} {5}},\
  \bibinfo {pages} {343} (\bibinfo {year} {2011})}\BibitemShut {NoStop}%
\bibitem [{\citenamefont {Franke-Arnold}(2017)}]{Franke2017}%
  \BibitemOpen
  \bibfield  {author} {\bibinfo {author} {\bibfnamefont {S.}~\bibnamefont
  {Franke-Arnold}},\ }\href {\doibase 10.1098/rsta.2015.0435} {\bibfield
  {journal} {\bibinfo  {journal} {Philosophical Transactions of the Royal
  Society A: Mathematical, Physical and Engineering Sciences}\ }\textbf
  {\bibinfo {volume} {375}},\ \bibinfo {pages} {20150435} (\bibinfo {year}
  {2017})}\BibitemShut {NoStop}%
\bibitem [{\citenamefont {Babiker}\ \emph {et~al.}(2018)\citenamefont
  {Babiker}, \citenamefont {Andrews},\ and\ \citenamefont
  {Lembessis}}]{Babiker2019}%
  \BibitemOpen
  \bibfield  {author} {\bibinfo {author} {\bibfnamefont {M.}~\bibnamefont
  {Babiker}}, \bibinfo {author} {\bibfnamefont {D.~L.}\ \bibnamefont
  {Andrews}}, \ and\ \bibinfo {author} {\bibfnamefont {V.~E.}\ \bibnamefont
  {Lembessis}},\ }\href {\doibase 10.1088/2040-8986/aaed14} {\bibfield
  {journal} {\bibinfo  {journal} {Journal of Optics}\ }\textbf {\bibinfo
  {volume} {21}},\ \bibinfo {pages} {013001} (\bibinfo {year}
  {2018})}\BibitemShut {NoStop}%
\bibitem [{\citenamefont {Kuga}\ \emph {et~al.}(1997)\citenamefont {Kuga},
  \citenamefont {Torii}, \citenamefont {Shiokawa}, \citenamefont {Hirano},
  \citenamefont {Shimizu},\ and\ \citenamefont {Sasada}}]{Kuga1997}%
  \BibitemOpen
  \bibfield  {author} {\bibinfo {author} {\bibfnamefont {T.}~\bibnamefont
  {Kuga}}, \bibinfo {author} {\bibfnamefont {Y.}~\bibnamefont {Torii}},
  \bibinfo {author} {\bibfnamefont {N.}~\bibnamefont {Shiokawa}}, \bibinfo
  {author} {\bibfnamefont {T.}~\bibnamefont {Hirano}}, \bibinfo {author}
  {\bibfnamefont {Y.}~\bibnamefont {Shimizu}}, \ and\ \bibinfo {author}
  {\bibfnamefont {H.}~\bibnamefont {Sasada}},\ }\href {\doibase
  10.1103/physrevlett.78.4713} {\bibfield  {journal} {\bibinfo  {journal}
  {Physical Review Letters}\ }\textbf {\bibinfo {volume} {78}},\ \bibinfo
  {pages} {4713} (\bibinfo {year} {1997})}\BibitemShut {NoStop}%
\bibitem [{\citenamefont {Chaloupka}\ \emph {et~al.}(1997)\citenamefont
  {Chaloupka}, \citenamefont {Fisher}, \citenamefont {Kessler},\ and\
  \citenamefont {Meyerhofer}}]{Chaloupka97}%
  \BibitemOpen
  \bibfield  {author} {\bibinfo {author} {\bibfnamefont {J.~L.}\ \bibnamefont
  {Chaloupka}}, \bibinfo {author} {\bibfnamefont {Y.}~\bibnamefont {Fisher}},
  \bibinfo {author} {\bibfnamefont {T.~J.}\ \bibnamefont {Kessler}}, \ and\
  \bibinfo {author} {\bibfnamefont {D.~D.}\ \bibnamefont {Meyerhofer}},\ }\href
  {\doibase 10.1364/ol.22.001021} {\bibfield  {journal} {\bibinfo  {journal}
  {Optics Letters}\ }\textbf {\bibinfo {volume} {22}},\ \bibinfo {pages} {1021}
  (\bibinfo {year} {1997})}\BibitemShut {NoStop}%
\bibitem [{\citenamefont {Ozeri}\ \emph {et~al.}(1999)\citenamefont {Ozeri},
  \citenamefont {Khaykovich},\ and\ \citenamefont {Davidson}}]{Ozeri99}%
  \BibitemOpen
  \bibfield  {author} {\bibinfo {author} {\bibfnamefont {R.}~\bibnamefont
  {Ozeri}}, \bibinfo {author} {\bibfnamefont {L.}~\bibnamefont {Khaykovich}}, \
  and\ \bibinfo {author} {\bibfnamefont {N.}~\bibnamefont {Davidson}},\ }\href
  {\doibase 10.1103/physreva.59.r1750} {\bibfield  {journal} {\bibinfo
  {journal} {Physical Review A}\ }\textbf {\bibinfo {volume} {59}},\ \bibinfo
  {pages} {R1750} (\bibinfo {year} {1999})}\BibitemShut {NoStop}%
\bibitem [{\citenamefont {Arlt}\ and\ \citenamefont
  {Padgett}(2000)}]{Arlt2000}%
  \BibitemOpen
  \bibfield  {author} {\bibinfo {author} {\bibfnamefont {J.}~\bibnamefont
  {Arlt}}\ and\ \bibinfo {author} {\bibfnamefont {M.~J.}\ \bibnamefont
  {Padgett}},\ }\href {\doibase 10.1364/ol.25.000191} {\bibfield  {journal}
  {\bibinfo  {journal} {Optics Letters}\ }\textbf {\bibinfo {volume} {25}},\
  \bibinfo {pages} {191} (\bibinfo {year} {2000})}\BibitemShut {NoStop}%
\bibitem [{\citenamefont {Arnold}(2012)}]{Arnold2012}%
  \BibitemOpen
  \bibfield  {author} {\bibinfo {author} {\bibfnamefont {A.~S.}\ \bibnamefont
  {Arnold}},\ }\href {\doibase 10.1364/ol.37.002505} {\bibfield  {journal}
  {\bibinfo  {journal} {Optics Letters}\ }\textbf {\bibinfo {volume} {37}},\
  \bibinfo {pages} {2505} (\bibinfo {year} {2012})}\BibitemShut {NoStop}%
\bibitem [{\citenamefont {Xu}\ \emph {et~al.}(2010)\citenamefont {Xu},
  \citenamefont {He}, \citenamefont {Wang},\ and\ \citenamefont
  {Zhan}}]{Xu2010}%
  \BibitemOpen
  \bibfield  {author} {\bibinfo {author} {\bibfnamefont {P.}~\bibnamefont
  {Xu}}, \bibinfo {author} {\bibfnamefont {X.}~\bibnamefont {He}}, \bibinfo
  {author} {\bibfnamefont {J.}~\bibnamefont {Wang}}, \ and\ \bibinfo {author}
  {\bibfnamefont {M.}~\bibnamefont {Zhan}},\ }\href {\doibase
  10.1364/ol.35.002164} {\bibfield  {journal} {\bibinfo  {journal} {Optics
  Letters}\ }\textbf {\bibinfo {volume} {35}},\ \bibinfo {pages} {2164}
  (\bibinfo {year} {2010})}\BibitemShut {NoStop}%
\bibitem [{\citenamefont {Barredo}\ \emph {et~al.}(2020)\citenamefont
  {Barredo}, \citenamefont {Lienhard}, \citenamefont {Scholl}, \citenamefont
  {de~L{\'e}s{\'e}leuc}, \citenamefont {Boulier}, \citenamefont {Browaeys},\
  and\ \citenamefont {Lahaye}}]{Barredo2020}%
  \BibitemOpen
  \bibfield  {author} {\bibinfo {author} {\bibfnamefont {D.}~\bibnamefont
  {Barredo}}, \bibinfo {author} {\bibfnamefont {V.}~\bibnamefont {Lienhard}},
  \bibinfo {author} {\bibfnamefont {P.}~\bibnamefont {Scholl}}, \bibinfo
  {author} {\bibfnamefont {S.}~\bibnamefont {de~L{\'e}s{\'e}leuc}}, \bibinfo
  {author} {\bibfnamefont {T.}~\bibnamefont {Boulier}}, \bibinfo {author}
  {\bibfnamefont {A.}~\bibnamefont {Browaeys}}, \ and\ \bibinfo {author}
  {\bibfnamefont {T.}~\bibnamefont {Lahaye}},\ }\href {\doibase
  10.1103/physrevlett.124.023201} {\bibfield  {journal} {\bibinfo  {journal}
  {Physical Review Letters}\ }\textbf {\bibinfo {volume} {124}},\ \bibinfo
  {pages} {023201} (\bibinfo {year} {2020})}\BibitemShut {NoStop}%
\bibitem [{\citenamefont {Hell}\ and\ \citenamefont {Stelzer}(1992)}]{Hell92}%
  \BibitemOpen
  \bibfield  {author} {\bibinfo {author} {\bibfnamefont {S.}~\bibnamefont
  {Hell}}\ and\ \bibinfo {author} {\bibfnamefont {E.~H.~K.}\ \bibnamefont
  {Stelzer}},\ }\href {\doibase 10.1364/josaa.9.002159} {\bibfield  {journal}
  {\bibinfo  {journal} {Journal of the Optical Society of America A}\ }\textbf
  {\bibinfo {volume} {9}},\ \bibinfo {pages} {2159} (\bibinfo {year}
  {1992})}\BibitemShut {NoStop}%
\bibitem [{\citenamefont {Bokor}\ and\ \citenamefont
  {Davidson}(2004)}]{Bokor2004}%
  \BibitemOpen
  \bibfield  {author} {\bibinfo {author} {\bibfnamefont {N.}~\bibnamefont
  {Bokor}}\ and\ \bibinfo {author} {\bibfnamefont {N.}~\bibnamefont
  {Davidson}},\ }\href {\doibase 10.1364/ol.29.001968} {\bibfield  {journal}
  {\bibinfo  {journal} {Optics Letters}\ }\textbf {\bibinfo {volume} {29}},\
  \bibinfo {pages} {1968} (\bibinfo {year} {2004})}\BibitemShut {NoStop}%
\bibitem [{\citenamefont {Boyd}(1980)}]{Boyd80}%
  \BibitemOpen
  \bibfield  {author} {\bibinfo {author} {\bibfnamefont {R.~W.}\ \bibnamefont
  {Boyd}},\ }\href {\doibase 10.1364/josa.70.000877} {\bibfield  {journal}
  {\bibinfo  {journal} {Journal of the Optical Society of America}\ }\textbf
  {\bibinfo {volume} {70}},\ \bibinfo {pages} {877} (\bibinfo {year}
  {1980})}\BibitemShut {NoStop}%
\bibitem [{\citenamefont {Steuernagel}\ \emph {et~al.}(2005)\citenamefont
  {Steuernagel}, \citenamefont {Yao}, \citenamefont {O'Holleran},\ and\
  \citenamefont {Padgett}}]{Steuernagel2005}%
  \BibitemOpen
  \bibfield  {author} {\bibinfo {author} {\bibfnamefont {O.}~\bibnamefont
  {Steuernagel}}, \bibinfo {author} {\bibfnamefont {E.}~\bibnamefont {Yao}},
  \bibinfo {author} {\bibfnamefont {K.}~\bibnamefont {O'Holleran}}, \ and\
  \bibinfo {author} {\bibfnamefont {M.}~\bibnamefont {Padgett}},\ }\href
  {\doibase 10.1080/09500340500347121} {\bibfield  {journal} {\bibinfo
  {journal} {Journal of Modern Optics}\ }\textbf {\bibinfo {volume} {52}},\
  \bibinfo {pages} {2713} (\bibinfo {year} {2005})}\BibitemShut {NoStop}%
\bibitem [{\citenamefont {Birr}\ \emph {et~al.}(2017)\citenamefont {Birr},
  \citenamefont {Fischer}, \citenamefont {Evlyukhin}, \citenamefont {Zywietz},
  \citenamefont {Chichkov},\ and\ \citenamefont {Reinhardt}}]{Birr2017}%
  \BibitemOpen
  \bibfield  {author} {\bibinfo {author} {\bibfnamefont {T.}~\bibnamefont
  {Birr}}, \bibinfo {author} {\bibfnamefont {T.}~\bibnamefont {Fischer}},
  \bibinfo {author} {\bibfnamefont {A.~B.}\ \bibnamefont {Evlyukhin}}, \bibinfo
  {author} {\bibfnamefont {U.}~\bibnamefont {Zywietz}}, \bibinfo {author}
  {\bibfnamefont {B.~N.}\ \bibnamefont {Chichkov}}, \ and\ \bibinfo {author}
  {\bibfnamefont {C.}~\bibnamefont {Reinhardt}},\ }\href {\doibase
  10.1021/acsphotonics.6b00999} {\bibfield  {journal} {\bibinfo  {journal} {ACS
  Photonics}\ }\textbf {\bibinfo {volume} {4}},\ \bibinfo {pages} {905}
  (\bibinfo {year} {2017})}\BibitemShut {NoStop}%
\bibitem [{\citenamefont {Isenhower}\ \emph {et~al.}(2009)\citenamefont
  {Isenhower}, \citenamefont {Williams}, \citenamefont {Dally},\ and\
  \citenamefont {Saffman}}]{Isenhower2009}%
  \BibitemOpen
  \bibfield  {author} {\bibinfo {author} {\bibfnamefont {L.}~\bibnamefont
  {Isenhower}}, \bibinfo {author} {\bibfnamefont {W.}~\bibnamefont {Williams}},
  \bibinfo {author} {\bibfnamefont {A.}~\bibnamefont {Dally}}, \ and\ \bibinfo
  {author} {\bibfnamefont {M.}~\bibnamefont {Saffman}},\ }\href {\doibase
  10.1364/ol.34.001159} {\bibfield  {journal} {\bibinfo  {journal} {Optics
  Letters}\ }\textbf {\bibinfo {volume} {34}},\ \bibinfo {pages} {1159}
  (\bibinfo {year} {2009})}\BibitemShut {NoStop}%
\bibitem [{\citenamefont {Whiting}\ \emph {et~al.}(2003)\citenamefont
  {Whiting}, \citenamefont {Abouraddy}, \citenamefont {Saleh}, \citenamefont
  {Teich},\ and\ \citenamefont {Fourkas}}]{Teich2003}%
  \BibitemOpen
  \bibfield  {author} {\bibinfo {author} {\bibfnamefont {A.}~\bibnamefont
  {Whiting}}, \bibinfo {author} {\bibfnamefont {A.}~\bibnamefont {Abouraddy}},
  \bibinfo {author} {\bibfnamefont {B.}~\bibnamefont {Saleh}}, \bibinfo
  {author} {\bibfnamefont {M.}~\bibnamefont {Teich}}, \ and\ \bibinfo {author}
  {\bibfnamefont {J.}~\bibnamefont {Fourkas}},\ }\href {\doibase
  10.1364/oe.11.001714} {\bibfield  {journal} {\bibinfo  {journal} {Optics
  Express}\ }\textbf {\bibinfo {volume} {11}},\ \bibinfo {pages} {1714}
  (\bibinfo {year} {2003})}\BibitemShut {NoStop}%
\bibitem [{\citenamefont {Freegarde}\ and\ \citenamefont
  {Dholakia}(2002)}]{Freegarde2002}%
  \BibitemOpen
  \bibfield  {author} {\bibinfo {author} {\bibfnamefont {T.}~\bibnamefont
  {Freegarde}}\ and\ \bibinfo {author} {\bibfnamefont {K.}~\bibnamefont
  {Dholakia}},\ }\href {\doibase 10.1103/physreva.66.013413} {\bibfield
  {journal} {\bibinfo  {journal} {Physical Review A}\ }\textbf {\bibinfo
  {volume} {66}},\ \bibinfo {pages} {013413} (\bibinfo {year}
  {2002})}\BibitemShut {NoStop}%
\bibitem [{\citenamefont {Chang}\ \emph {et~al.}(2018)\citenamefont {Chang},
  \citenamefont {Douglas}, \citenamefont {Gonz\'alez-Tudela}, \citenamefont
  {Hung},\ and\ \citenamefont {Kimble}}]{Kimble:18}%
  \BibitemOpen
  \bibfield  {author} {\bibinfo {author} {\bibfnamefont {D.~E.}\ \bibnamefont
  {Chang}}, \bibinfo {author} {\bibfnamefont {J.~S.}\ \bibnamefont {Douglas}},
  \bibinfo {author} {\bibfnamefont {A.}~\bibnamefont {Gonz\'alez-Tudela}},
  \bibinfo {author} {\bibfnamefont {C.-L.}\ \bibnamefont {Hung}}, \ and\
  \bibinfo {author} {\bibfnamefont {H.~J.}\ \bibnamefont {Kimble}},\ }\href
  {\doibase 10.1103/RevModPhys.90.031002} {\bibfield  {journal} {\bibinfo
  {journal} {Rev. Mod. Phys.}\ }\textbf {\bibinfo {volume} {90}},\ \bibinfo
  {pages} {031002} (\bibinfo {year} {2018})}\BibitemShut {NoStop}%
\bibitem [{\citenamefont {Allen}\ \emph {et~al.}(1992)\citenamefont {Allen},
  \citenamefont {Beijersbergen}, \citenamefont {Spreeuw},\ and\ \citenamefont
  {Woerdman}}]{Allen1992}%
  \BibitemOpen
  \bibfield  {author} {\bibinfo {author} {\bibfnamefont {L.}~\bibnamefont
  {Allen}}, \bibinfo {author} {\bibfnamefont {M.~W.}\ \bibnamefont
  {Beijersbergen}}, \bibinfo {author} {\bibfnamefont {R.~J.~C.}\ \bibnamefont
  {Spreeuw}}, \ and\ \bibinfo {author} {\bibfnamefont {J.~P.}\ \bibnamefont
  {Woerdman}},\ }\href {\doibase 10.1103/physreva.45.8185} {\bibfield
  {journal} {\bibinfo  {journal} {Physical Review A}\ }\textbf {\bibinfo
  {volume} {45}},\ \bibinfo {pages} {8185} (\bibinfo {year}
  {1992})}\BibitemShut {NoStop}%
\bibitem [{\citenamefont {Siegman}(1986)}]{bookSiegman}%
  \BibitemOpen
  \bibfield  {author} {\bibinfo {author} {\bibfnamefont {A.~E.}\ \bibnamefont
  {Siegman}},\ }\href@noop {} {\emph {\bibinfo {title} {Lasers}}}\ (\bibinfo
  {publisher} {Oxford University, Oxford},\ \bibinfo {year} {1986})\BibitemShut
  {NoStop}%
\bibitem [{SM(2020)}]{SM}%
  \BibitemOpen
  \href@noop {} {\enquote {\bibinfo {title} {Refer to accompanying supplemental
  material.}}\ } (\bibinfo {year} {2020})\BibitemShut {NoStop}%
\bibitem [{\citenamefont {Phillips}\ and\ \citenamefont
  {Andrews}(1983)}]{Phillips1983}%
  \BibitemOpen
  \bibfield  {author} {\bibinfo {author} {\bibfnamefont {R.~L.}\ \bibnamefont
  {Phillips}}\ and\ \bibinfo {author} {\bibfnamefont {L.~C.}\ \bibnamefont
  {Andrews}},\ }\href {\doibase 10.1364/ao.22.000643} {\bibfield  {journal}
  {\bibinfo  {journal} {Applied Optics}\ }\textbf {\bibinfo {volume} {22}},\
  \bibinfo {pages} {643} (\bibinfo {year} {1983})}\BibitemShut {NoStop}%
\bibitem [{\citenamefont {Ando}\ \emph {et~al.}(2008)\citenamefont {Ando},
  \citenamefont {Ohtake}, \citenamefont {Matsumoto}, \citenamefont {Inoue},\
  and\ \citenamefont {Fukuchi}}]{Ando2009}%
  \BibitemOpen
  \bibfield  {author} {\bibinfo {author} {\bibfnamefont {T.}~\bibnamefont
  {Ando}}, \bibinfo {author} {\bibfnamefont {Y.}~\bibnamefont {Ohtake}},
  \bibinfo {author} {\bibfnamefont {N.}~\bibnamefont {Matsumoto}}, \bibinfo
  {author} {\bibfnamefont {T.}~\bibnamefont {Inoue}}, \ and\ \bibinfo {author}
  {\bibfnamefont {N.}~\bibnamefont {Fukuchi}},\ }\href {\doibase
  10.1364/ol.34.000034} {\bibfield  {journal} {\bibinfo  {journal} {Optics
  Letters}\ }\textbf {\bibinfo {volume} {34}},\ \bibinfo {pages} {34} (\bibinfo
  {year} {2008})}\BibitemShut {NoStop}%
\bibitem [{\citenamefont {Arlt}\ \emph {et~al.}(1998)\citenamefont {Arlt},
  \citenamefont {Dholakia}, \citenamefont {Allen},\ and\ \citenamefont
  {Padgett}}]{Arlt1998}%
  \BibitemOpen
  \bibfield  {author} {\bibinfo {author} {\bibfnamefont {J.}~\bibnamefont
  {Arlt}}, \bibinfo {author} {\bibfnamefont {K.}~\bibnamefont {Dholakia}},
  \bibinfo {author} {\bibfnamefont {L.}~\bibnamefont {Allen}}, \ and\ \bibinfo
  {author} {\bibfnamefont {M.~J.}\ \bibnamefont {Padgett}},\ }\href {\doibase
  10.1080/09500349808230913} {\bibfield  {journal} {\bibinfo  {journal}
  {Journal of Modern Optics}\ }\textbf {\bibinfo {volume} {45}},\ \bibinfo
  {pages} {1231} (\bibinfo {year} {1998})}\BibitemShut {NoStop}%
\bibitem [{\citenamefont {Davis}\ \emph {et~al.}(1999)\citenamefont {Davis},
  \citenamefont {Cottrell}, \citenamefont {Campos}, \citenamefont {Yzuel},\
  and\ \citenamefont {Moreno}}]{Davis1999}%
  \BibitemOpen
  \bibfield  {author} {\bibinfo {author} {\bibfnamefont {J.~A.}\ \bibnamefont
  {Davis}}, \bibinfo {author} {\bibfnamefont {D.~M.}\ \bibnamefont {Cottrell}},
  \bibinfo {author} {\bibfnamefont {J.}~\bibnamefont {Campos}}, \bibinfo
  {author} {\bibfnamefont {M.~J.}\ \bibnamefont {Yzuel}}, \ and\ \bibinfo
  {author} {\bibfnamefont {I.}~\bibnamefont {Moreno}},\ }\href {\doibase
  10.1364/ao.38.005004} {\bibfield  {journal} {\bibinfo  {journal} {Applied
  Optics}\ }\textbf {\bibinfo {volume} {38}},\ \bibinfo {pages} {5004}
  (\bibinfo {year} {1999})}\BibitemShut {NoStop}%
\bibitem [{\citenamefont {Bolduc}\ \emph {et~al.}(2013)\citenamefont {Bolduc},
  \citenamefont {Bent}, \citenamefont {Santamato}, \citenamefont {Karimi},\
  and\ \citenamefont {Boyd}}]{Bolduc2013}%
  \BibitemOpen
  \bibfield  {author} {\bibinfo {author} {\bibfnamefont {E.}~\bibnamefont
  {Bolduc}}, \bibinfo {author} {\bibfnamefont {N.}~\bibnamefont {Bent}},
  \bibinfo {author} {\bibfnamefont {E.}~\bibnamefont {Santamato}}, \bibinfo
  {author} {\bibfnamefont {E.}~\bibnamefont {Karimi}}, \ and\ \bibinfo {author}
  {\bibfnamefont {R.~W.}\ \bibnamefont {Boyd}},\ }\href {\doibase
  10.1364/ol.38.003546} {\bibfield  {journal} {\bibinfo  {journal} {Optics
  Letters}\ }\textbf {\bibinfo {volume} {38}},\ \bibinfo {pages} {3546}
  (\bibinfo {year} {2013})}\BibitemShut {NoStop}%
\bibitem [{\citenamefont {Richards}\ and\ \citenamefont
  {Wolf}(1959)}]{Richards1959}%
  \BibitemOpen
  \bibfield  {author} {\bibinfo {author} {\bibfnamefont {B.}~\bibnamefont
  {Richards}}\ and\ \bibinfo {author} {\bibfnamefont {E.}~\bibnamefont
  {Wolf}},\ }\href {\doibase 10.1098/rspa.1959.0200} {\bibfield  {journal}
  {\bibinfo  {journal} {Proceedings of the Royal Society of London. Series A.
  Mathematical and Physical Sciences}\ }\textbf {\bibinfo {volume} {253}},\
  \bibinfo {pages} {358} (\bibinfo {year} {1959})}\BibitemShut {NoStop}%
\bibitem [{\citenamefont {Novotny}\ and\ \citenamefont
  {Hecht}(2006)}]{NovotnyBook}%
  \BibitemOpen
  \bibfield  {author} {\bibinfo {author} {\bibfnamefont {L.}~\bibnamefont
  {Novotny}}\ and\ \bibinfo {author} {\bibfnamefont {B.}~\bibnamefont
  {Hecht}},\ }\href@noop {} {\emph {\bibinfo {title} {Principles of
  Nano-Optics}}}\ (\bibinfo  {publisher} {Cambridge University Press,
  Cambridge, UK},\ \bibinfo {year} {2006})\BibitemShut {NoStop}%
\bibitem [{\citenamefont {Luan}(2020)}]{Luan_thesis}%
  \BibitemOpen
  \bibfield  {author} {\bibinfo {author} {\bibfnamefont {X.}~\bibnamefont
  {Luan}},\ }\emph {\bibinfo {title} {Towards Atom Assembly on Nanophotonic
  Structures with Optical Tweezers}},\ \href {\doibase 10.7907/08q5-0w11}
  {Ph.D. thesis},\ \bibinfo  {school} {California Institute of Technology}
  (\bibinfo {year} {2020})\BibitemShut {NoStop}%
\bibitem [{\citenamefont {Haddadi}\ \emph {et~al.}(2015)\citenamefont
  {Haddadi}, \citenamefont {Louhibi}, \citenamefont {Hasnaoui}, \citenamefont
  {Harfouche},\ and\ \citenamefont {A\"{i}t-Ameur}}]{Haddadi2015}%
  \BibitemOpen
  \bibfield  {author} {\bibinfo {author} {\bibfnamefont {S.}~\bibnamefont
  {Haddadi}}, \bibinfo {author} {\bibfnamefont {D.}~\bibnamefont {Louhibi}},
  \bibinfo {author} {\bibfnamefont {A.}~\bibnamefont {Hasnaoui}}, \bibinfo
  {author} {\bibfnamefont {A.}~\bibnamefont {Harfouche}}, \ and\ \bibinfo
  {author} {\bibfnamefont {K.}~\bibnamefont {A\"{i}t-Ameur}},\ }\href {\doibase
  10.1088/1054-660x/25/12/125002} {\bibfield  {journal} {\bibinfo  {journal}
  {Laser Physics}\ }\textbf {\bibinfo {volume} {25}},\ \bibinfo {pages}
  {125002} (\bibinfo {year} {2015})}\BibitemShut {NoStop}%
\bibitem [{\citenamefont {Kuhr}\ \emph {et~al.}(2005)\citenamefont {Kuhr},
  \citenamefont {Alt}, \citenamefont {Schrader}, \citenamefont {Dotsenko},
  \citenamefont {Miroshnychenko}, \citenamefont {Rauschenbeutel},\ and\
  \citenamefont {Meschede}}]{Kuhr2005}%
  \BibitemOpen
  \bibfield  {author} {\bibinfo {author} {\bibfnamefont {S.}~\bibnamefont
  {Kuhr}}, \bibinfo {author} {\bibfnamefont {W.}~\bibnamefont {Alt}}, \bibinfo
  {author} {\bibfnamefont {D.}~\bibnamefont {Schrader}}, \bibinfo {author}
  {\bibfnamefont {I.}~\bibnamefont {Dotsenko}}, \bibinfo {author}
  {\bibfnamefont {Y.}~\bibnamefont {Miroshnychenko}}, \bibinfo {author}
  {\bibfnamefont {A.}~\bibnamefont {Rauschenbeutel}}, \ and\ \bibinfo {author}
  {\bibfnamefont {D.}~\bibnamefont {Meschede}},\ }\href {\doibase
  10.1103/physreva.72.023406} {\bibfield  {journal} {\bibinfo  {journal}
  {Physical Review A}\ }\textbf {\bibinfo {volume} {72}},\ \bibinfo {pages}
  {023406} (\bibinfo {year} {2005})}\BibitemShut {NoStop}%
\bibitem [{\citenamefont {Thompson}\ \emph
  {et~al.}(2013{\natexlab{a}})\citenamefont {Thompson}, \citenamefont {Tiecke},
  \citenamefont {Zibrov}, \citenamefont {Vuleti\ifmmode~\acute{c}\else
  \'{c}\fi{}},\ and\ \citenamefont {Lukin}}]{ThompsonPRL:13}%
  \BibitemOpen
  \bibfield  {author} {\bibinfo {author} {\bibfnamefont {J.~D.}\ \bibnamefont
  {Thompson}}, \bibinfo {author} {\bibfnamefont {T.~G.}\ \bibnamefont
  {Tiecke}}, \bibinfo {author} {\bibfnamefont {A.~S.}\ \bibnamefont {Zibrov}},
  \bibinfo {author} {\bibfnamefont {V.}~\bibnamefont
  {Vuleti\ifmmode~\acute{c}\else \'{c}\fi{}}}, \ and\ \bibinfo {author}
  {\bibfnamefont {M.~D.}\ \bibnamefont {Lukin}},\ }\href {\doibase
  10.1103/PhysRevLett.110.133001} {\bibfield  {journal} {\bibinfo  {journal}
  {Phys. Rev. Lett.}\ }\textbf {\bibinfo {volume} {110}},\ \bibinfo {pages}
  {133001} (\bibinfo {year} {2013}{\natexlab{a}})}\BibitemShut {NoStop}%
\bibitem [{\citenamefont {Goban}\ \emph {et~al.}(2012)\citenamefont {Goban},
  \citenamefont {Choi}, \citenamefont {Alton}, \citenamefont {Ding},
  \citenamefont {Lacro\^ute}, \citenamefont {Pototschnig}, \citenamefont
  {Thiele}, \citenamefont {Stern},\ and\ \citenamefont
  {Kimble}}]{Goban2012Demonstration}%
  \BibitemOpen
  \bibfield  {author} {\bibinfo {author} {\bibfnamefont {A.}~\bibnamefont
  {Goban}}, \bibinfo {author} {\bibfnamefont {K.~S.}\ \bibnamefont {Choi}},
  \bibinfo {author} {\bibfnamefont {D.~J.}\ \bibnamefont {Alton}}, \bibinfo
  {author} {\bibfnamefont {D.}~\bibnamefont {Ding}}, \bibinfo {author}
  {\bibfnamefont {C.}~\bibnamefont {Lacro\^ute}}, \bibinfo {author}
  {\bibfnamefont {M.}~\bibnamefont {Pototschnig}}, \bibinfo {author}
  {\bibfnamefont {T.}~\bibnamefont {Thiele}}, \bibinfo {author} {\bibfnamefont
  {N.~P.}\ \bibnamefont {Stern}}, \ and\ \bibinfo {author} {\bibfnamefont
  {H.~J.}\ \bibnamefont {Kimble}},\ }\href {\doibase
  10.1103/PhysRevLett.109.033603} {\bibfield  {journal} {\bibinfo  {journal}
  {Phys. Rev. Lett.}\ }\textbf {\bibinfo {volume} {109}},\ \bibinfo {pages}
  {033603} (\bibinfo {year} {2012})}\BibitemShut {NoStop}%
\bibitem [{\citenamefont {H\"ummer}\ \emph {et~al.}(2019)\citenamefont
  {H\"ummer}, \citenamefont {Schneeweiss}, \citenamefont {Rauschenbeutel},\
  and\ \citenamefont {Romero-Isart}}]{hummer19}%
  \BibitemOpen
  \bibfield  {author} {\bibinfo {author} {\bibfnamefont {D.}~\bibnamefont
  {H\"ummer}}, \bibinfo {author} {\bibfnamefont {P.}~\bibnamefont
  {Schneeweiss}}, \bibinfo {author} {\bibfnamefont {A.}~\bibnamefont
  {Rauschenbeutel}}, \ and\ \bibinfo {author} {\bibfnamefont {O.}~\bibnamefont
  {Romero-Isart}},\ }\href {\doibase 10.1103/PhysRevX.9.041034} {\bibfield
  {journal} {\bibinfo  {journal} {Phys. Rev. X}\ }\textbf {\bibinfo {volume}
  {9}},\ \bibinfo {pages} {041034} (\bibinfo {year} {2019})}\BibitemShut
  {NoStop}%
\bibitem [{\citenamefont {Durnin}\ \emph {et~al.}(1987)\citenamefont {Durnin},
  \citenamefont {Miceli},\ and\ \citenamefont {Eberly}}]{Durnin:87:Bessel}%
  \BibitemOpen
  \bibfield  {author} {\bibinfo {author} {\bibfnamefont {J.}~\bibnamefont
  {Durnin}}, \bibinfo {author} {\bibfnamefont {J.~J.}\ \bibnamefont {Miceli}},
  \ and\ \bibinfo {author} {\bibfnamefont {J.~H.}\ \bibnamefont {Eberly}},\
  }\href {\doibase 10.1103/PhysRevLett.58.1499} {\bibfield  {journal} {\bibinfo
   {journal} {Phys. Rev. Lett.}\ }\textbf {\bibinfo {volume} {58}},\ \bibinfo
  {pages} {1499} (\bibinfo {year} {1987})}\BibitemShut {NoStop}%
\bibitem [{\citenamefont {Tiecke}\ \emph {et~al.}(2014)\citenamefont {Tiecke},
  \citenamefont {Thompson}, \citenamefont {de~Leon}, \citenamefont {Liu},
  \citenamefont {Vuleti{\'c}},\ and\ \citenamefont {Lukin}}]{Tiecke:14}%
  \BibitemOpen
  \bibfield  {author} {\bibinfo {author} {\bibfnamefont {T.}~\bibnamefont
  {Tiecke}}, \bibinfo {author} {\bibfnamefont {J.~D.}\ \bibnamefont
  {Thompson}}, \bibinfo {author} {\bibfnamefont {N.~P.}\ \bibnamefont
  {de~Leon}}, \bibinfo {author} {\bibfnamefont {L.}~\bibnamefont {Liu}},
  \bibinfo {author} {\bibfnamefont {V.}~\bibnamefont {Vuleti{\'c}}}, \ and\
  \bibinfo {author} {\bibfnamefont {M.~D.}\ \bibnamefont {Lukin}},\ }\href
  {\doibase 10.1038/nature13188} {\bibfield  {journal} {\bibinfo  {journal}
  {Nature}\ }\textbf {\bibinfo {volume} {508}},\ \bibinfo {pages} {241}
  (\bibinfo {year} {2014})}\BibitemShut {NoStop}%
\bibitem [{\citenamefont {Hood}\ \emph {et~al.}(2016)\citenamefont {Hood},
  \citenamefont {Goban}, \citenamefont {Asenjo-Garcia}, \citenamefont {Lu},
  \citenamefont {Yu}, \citenamefont {Chang},\ and\ \citenamefont
  {Kimble}}]{Hood:16}%
  \BibitemOpen
  \bibfield  {author} {\bibinfo {author} {\bibfnamefont {J.~D.}\ \bibnamefont
  {Hood}}, \bibinfo {author} {\bibfnamefont {A.}~\bibnamefont {Goban}},
  \bibinfo {author} {\bibfnamefont {A.}~\bibnamefont {Asenjo-Garcia}}, \bibinfo
  {author} {\bibfnamefont {M.}~\bibnamefont {Lu}}, \bibinfo {author}
  {\bibfnamefont {S.-P.}\ \bibnamefont {Yu}}, \bibinfo {author} {\bibfnamefont
  {D.~E.}\ \bibnamefont {Chang}}, \ and\ \bibinfo {author} {\bibfnamefont
  {H.~J.}\ \bibnamefont {Kimble}},\ }\href {\doibase 10.1073/pnas.1603788113}
  {\bibfield  {journal} {\bibinfo  {journal} {Proc. Natl. Acad. Sci. U.S.A.}\
  }\textbf {\bibinfo {volume} {113}},\ \bibinfo {pages} {10507} (\bibinfo
  {year} {2016})}\BibitemShut {NoStop}%
\bibitem [{\citenamefont {Burgers}\ \emph {et~al.}(2019)\citenamefont
  {Burgers}, \citenamefont {Peng}, \citenamefont {Muniz}, \citenamefont
  {McClung}, \citenamefont {Martin},\ and\ \citenamefont
  {Kimble}}]{Burgers:19}%
  \BibitemOpen
  \bibfield  {author} {\bibinfo {author} {\bibfnamefont {A.~P.}\ \bibnamefont
  {Burgers}}, \bibinfo {author} {\bibfnamefont {L.~S.}\ \bibnamefont {Peng}},
  \bibinfo {author} {\bibfnamefont {J.~A.}\ \bibnamefont {Muniz}}, \bibinfo
  {author} {\bibfnamefont {A.~C.}\ \bibnamefont {McClung}}, \bibinfo {author}
  {\bibfnamefont {M.~J.}\ \bibnamefont {Martin}}, \ and\ \bibinfo {author}
  {\bibfnamefont {H.~J.}\ \bibnamefont {Kimble}},\ }\href {\doibase
  10.1073/pnas.1817249115} {\bibfield  {journal} {\bibinfo  {journal} {Proc.
  Natl. Acad. Sci. U.S.A.}\ }\textbf {\bibinfo {volume} {116}},\ \bibinfo
  {pages} {456} (\bibinfo {year} {2019})},\ \Eprint
  {http://arxiv.org/abs/https://www.pnas.org/content/116/2/456.full.pdf}
  {https://www.pnas.org/content/116/2/456.full.pdf} \BibitemShut {NoStop}%
\bibitem [{\citenamefont {Kim}\ \emph {et~al.}(2019)\citenamefont {Kim},
  \citenamefont {Chang}, \citenamefont {Fields}, \citenamefont {Chen},\ and\
  \citenamefont {Hung}}]{Kim:19}%
  \BibitemOpen
  \bibfield  {author} {\bibinfo {author} {\bibfnamefont {M.~E.}\ \bibnamefont
  {Kim}}, \bibinfo {author} {\bibfnamefont {T.-H.}\ \bibnamefont {Chang}},
  \bibinfo {author} {\bibfnamefont {B.~M.}\ \bibnamefont {Fields}}, \bibinfo
  {author} {\bibfnamefont {C.-A.}\ \bibnamefont {Chen}}, \ and\ \bibinfo
  {author} {\bibfnamefont {C.-L.}\ \bibnamefont {Hung}},\ }\href {\doibase
  10.1038/s41467-019-09635-7} {\bibfield  {journal} {\bibinfo  {journal}
  {Nature Communications}\ }\textbf {\bibinfo {volume} {10}},\ \bibinfo {pages}
  {1647} (\bibinfo {year} {2019})}\BibitemShut {NoStop}%
\bibitem [{\citenamefont {Thompson}\ \emph
  {et~al.}(2013{\natexlab{b}})\citenamefont {Thompson}, \citenamefont {Tiecke},
  \citenamefont {de~Leon}, \citenamefont {Feist}, \citenamefont {Akimov},
  \citenamefont {Gullans}, \citenamefont {Zibrov}, \citenamefont
  {Vuleti{\'c}},\ and\ \citenamefont {Lukin}}]{ThompsonSci:2013}%
  \BibitemOpen
  \bibfield  {author} {\bibinfo {author} {\bibfnamefont {J.~D.}\ \bibnamefont
  {Thompson}}, \bibinfo {author} {\bibfnamefont {T.}~\bibnamefont {Tiecke}},
  \bibinfo {author} {\bibfnamefont {N.~P.}\ \bibnamefont {de~Leon}}, \bibinfo
  {author} {\bibfnamefont {J.}~\bibnamefont {Feist}}, \bibinfo {author}
  {\bibfnamefont {A.}~\bibnamefont {Akimov}}, \bibinfo {author} {\bibfnamefont
  {M.}~\bibnamefont {Gullans}}, \bibinfo {author} {\bibfnamefont {A.~S.}\
  \bibnamefont {Zibrov}}, \bibinfo {author} {\bibfnamefont {V.}~\bibnamefont
  {Vuleti{\'c}}}, \ and\ \bibinfo {author} {\bibfnamefont {M.~D.}\ \bibnamefont
  {Lukin}},\ }\href {\doibase 10.1126/science.1237125} {\bibfield  {journal}
  {\bibinfo  {journal} {Science}\ }\textbf {\bibinfo {volume} {340}},\ \bibinfo
  {pages} {1202} (\bibinfo {year} {2013}{\natexlab{b}})}\BibitemShut {NoStop}%
\bibitem [{\citenamefont {Yu}\ \emph {et~al.}(2014)\citenamefont {Yu},
  \citenamefont {Hood}, \citenamefont {Muniz}, \citenamefont {Martin},
  \citenamefont {Norte}, \citenamefont {Hung}, \citenamefont {Meenehan},
  \citenamefont {Cohen}, \citenamefont {Painter},\ and\ \citenamefont
  {Kimble}}]{Yu:14}%
  \BibitemOpen
  \bibfield  {author} {\bibinfo {author} {\bibfnamefont {S.-P.}\ \bibnamefont
  {Yu}}, \bibinfo {author} {\bibfnamefont {J.~D.}\ \bibnamefont {Hood}},
  \bibinfo {author} {\bibfnamefont {J.~A.}\ \bibnamefont {Muniz}}, \bibinfo
  {author} {\bibfnamefont {M.~J.}\ \bibnamefont {Martin}}, \bibinfo {author}
  {\bibfnamefont {R.}~\bibnamefont {Norte}}, \bibinfo {author} {\bibfnamefont
  {C.-L.}\ \bibnamefont {Hung}}, \bibinfo {author} {\bibfnamefont {S.~M.}\
  \bibnamefont {Meenehan}}, \bibinfo {author} {\bibfnamefont {J.~D.}\
  \bibnamefont {Cohen}}, \bibinfo {author} {\bibfnamefont {O.}~\bibnamefont
  {Painter}}, \ and\ \bibinfo {author} {\bibfnamefont {H.~J.}\ \bibnamefont
  {Kimble}},\ }\href {\doibase 10.1063/1.4868975} {\bibfield  {journal}
  {\bibinfo  {journal} {Applied Physics Letters}\ }\textbf {\bibinfo {volume}
  {104}},\ \bibinfo {pages} {111103} (\bibinfo {year} {2014})},\ \Eprint
  {http://arxiv.org/abs/https://doi.org/10.1063/1.4868975}
  {https://doi.org/10.1063/1.4868975} \BibitemShut {NoStop}%
\bibitem [{\citenamefont {Yu}\ \emph {et~al.}(2019)\citenamefont {Yu},
  \citenamefont {Muniz}, \citenamefont {Hung},\ and\ \citenamefont
  {Kimble}}]{Yu2019}%
  \BibitemOpen
  \bibfield  {author} {\bibinfo {author} {\bibfnamefont {S.-P.}\ \bibnamefont
  {Yu}}, \bibinfo {author} {\bibfnamefont {J.~A.}\ \bibnamefont {Muniz}},
  \bibinfo {author} {\bibfnamefont {C.-L.}\ \bibnamefont {Hung}}, \ and\
  \bibinfo {author} {\bibfnamefont {H.~J.}\ \bibnamefont {Kimble}},\ }\href
  {\doibase 10.1073/pnas.1822110116} {\bibfield  {journal} {\bibinfo  {journal}
  {Proc. Natl. Acad. Sci. U.S.A.}\ }\textbf {\bibinfo {volume} {116}},\
  \bibinfo {pages} {12743} (\bibinfo {year} {2019})}\BibitemShut {NoStop}%
\bibitem [{\citenamefont {Gonz{\'a}lez-Tudela}\ \emph
  {et~al.}(2015)\citenamefont {Gonz{\'a}lez-Tudela}, \citenamefont {Hung},
  \citenamefont {Chang}, \citenamefont {Cirac},\ and\ \citenamefont
  {Kimble}}]{Tudela2015}%
  \BibitemOpen
  \bibfield  {author} {\bibinfo {author} {\bibfnamefont {A.}~\bibnamefont
  {Gonz{\'a}lez-Tudela}}, \bibinfo {author} {\bibfnamefont {C.-L.}\
  \bibnamefont {Hung}}, \bibinfo {author} {\bibfnamefont {D.~E.}\ \bibnamefont
  {Chang}}, \bibinfo {author} {\bibfnamefont {J.~I.}\ \bibnamefont {Cirac}}, \
  and\ \bibinfo {author} {\bibfnamefont {H.~J.}\ \bibnamefont {Kimble}},\
  }\href {\doibase 10.1038/nphoton.2015.54} {\bibfield  {journal} {\bibinfo
  {journal} {Nature Photonics}\ }\textbf {\bibinfo {volume} {9}},\ \bibinfo
  {pages} {320} (\bibinfo {year} {2015})}\BibitemShut {NoStop}%
\bibitem [{\citenamefont {B\'{e}guin}\ \emph {et~al.}(2020)\citenamefont
  {B\'{e}guin}, \citenamefont {Burgers}, \citenamefont {Luan}, \citenamefont
  {Qin}, \citenamefont {Yu},\ and\ \citenamefont {Kimble}}]{Beguin2020}%
  \BibitemOpen
  \bibfield  {author} {\bibinfo {author} {\bibfnamefont {J.-B.}\ \bibnamefont
  {B\'{e}guin}}, \bibinfo {author} {\bibfnamefont {A.~P.}\ \bibnamefont
  {Burgers}}, \bibinfo {author} {\bibfnamefont {X.}~\bibnamefont {Luan}},
  \bibinfo {author} {\bibfnamefont {Z.}~\bibnamefont {Qin}}, \bibinfo {author}
  {\bibfnamefont {S.~P.}\ \bibnamefont {Yu}}, \ and\ \bibinfo {author}
  {\bibfnamefont {H.~J.}\ \bibnamefont {Kimble}},\ }\href {\doibase
  10.1364/OPTICA.384408} {\bibfield  {journal} {\bibinfo  {journal} {Optica}\
  }\textbf {\bibinfo {volume} {7}},\ \bibinfo {pages} {1} (\bibinfo {year}
  {2020})}\BibitemShut {NoStop}%
\bibitem [{\citenamefont {Luan}\ \emph {et~al.}(2020)\citenamefont {Luan},
  \citenamefont {Béguin}, \citenamefont {Burgers}, \citenamefont {Qin},
  \citenamefont {Yu},\ and\ \citenamefont {Kimble}}]{Luan2020integration}%
  \BibitemOpen
  \bibfield  {author} {\bibinfo {author} {\bibfnamefont {X.}~\bibnamefont
  {Luan}}, \bibinfo {author} {\bibfnamefont {J.-B.}\ \bibnamefont {Béguin}},
  \bibinfo {author} {\bibfnamefont {A.~P.}\ \bibnamefont {Burgers}}, \bibinfo
  {author} {\bibfnamefont {Z.}~\bibnamefont {Qin}}, \bibinfo {author}
  {\bibfnamefont {S.-P.}\ \bibnamefont {Yu}}, \ and\ \bibinfo {author}
  {\bibfnamefont {H.~J.}\ \bibnamefont {Kimble}},\ }\href {\doibase
  10.1002/qute.202000008} {\bibfield  {journal} {\bibinfo  {journal} {Advanced
  Quantum Technologies}\ ,\ \bibinfo {pages} {2000008}} (\bibinfo {year}
  {2020})},\ \Eprint
  {http://arxiv.org/abs/https://onlinelibrary.wiley.com/doi/pdf/10.1002/qute.202000008}
  {https://onlinelibrary.wiley.com/doi/pdf/10.1002/qute.202000008} \BibitemShut
  {NoStop}%
\end{thebibliography}%

\pagebreak
\widetext
\begin{center}
\textbf{\large Supplemental Materials}
\end{center}
\setcounter{equation}{0}
\setcounter{figure}{0}
\setcounter{table}{0}
\renewcommand{\theequation}{S\arabic{equation}}
\renewcommand{\thefigure}{S\arabic{figure}}
\setcounter{section}{0}
\section{Field expression}

For convenience, the expression for the complex scalar amplitude function for conventional Laguerre-Gauss beams is reproduced here from~\cite{Allen1992,bookSiegman}
\begin{align}
u_{p}(r,z) = \sqrt{\frac{2}{\pi}}\frac{w_0}{w(z)}\exp\left(-\frac{r^2}{w(z)^2}\right)\exp\left(-ik\frac{r^2}{2R(z)}\right)\times L_p^0\left(\frac{2r^2}{w(z)^2}\right) \exp\left(i\Psi_p (z)\right),
\end{align}
 with $w_0$ the waist, $z_R=\pi w_0^2/\lambda$ the Rayleigh range, $R(z)=z(1+z_R^2/z^2)$ the radius of curvature, and $w(z)=w_0\sqrt{1+z^2/z_R^2}$ the waist at position $z$. The Gouy phase is given by \mbox{$\Psi_p (z)=(2p+1)\textrm{arctan}(z/z_R)$} and $L_p^0$ correponds to the associated Laguerre polynomial.

\section{Trap frequencies and dimensions}

\subsection{The paraxial limit}

The full widths at half maxima for the intensity distributions along $x,y,z$ in Fig. 1 of the main text are $\Delta x_0 = \Delta y_0 =  \SI{1.17}{\micro\meter}$ and $\Delta z_0 = \SI{6.28}{\micro\meter}$ for \mbox{$\vec{E}_{0}$}, and $\Delta x_\Sigma = \Delta y_\Sigma =  \SI{0.51}{\micro\meter}$ and $\Delta z_\Sigma = \SI{1.5}{\micro\meter}$ for \mbox{$\vec{E}_{\Sigma}$}. One metric for confinement of an atom of mass $M$ in an optical tweezer is the frequency of oscillation near the bottom of the tweezer's optical potential. Relative to the trap formed by \mbox{$\vec{E}_{0}$} in Fig. 1 of the main text, transverse and longitudinal trap frequencies in the paraxial limit for \mbox{$\vec{E}_{\Sigma}$} are increased as $\omega_x^{\Sigma} = \omega_x^0 \sqrt{5}$ and $\omega_z^{\Sigma} = \omega_z^0\sqrt{35/3}$, with trap depth $U_0$ for both \mbox{$\vec{E}_{0}$} and \mbox{$\vec{E}_{\Sigma}$}. For Cs atoms with $U_0 = 1$mK, wavelength $\lambda = \SI{1}{\micro\meter}$, and waist $w_0 = \SI{1}{\micro\meter}$, $\omega_x^{\Sigma} = 2\pi\times\SI{178}{\kilo\hertz}$ and $\omega_z^{\Sigma} = 2\pi\times\SI{61}{\kilo\hertz}$. Here, $\omega_x^0 = \sqrt{4U_0 /M w^2}$ and $\omega_z^0 = \sqrt{2U_0 /M z_R^2}$ are the transverse and longitudinal angular trap frequencies for an ideal Gaussian mode \mbox{$\vec{E}_{0}$} in the paraxial limit, with an accuracy better than $1\%$ as compared to the ground state trap frequencies obtained by numerical solution of the spatial Schr\"odinger equation.

\subsection{The vector theory}
\begin{figure}[htbp]
\centering
\includegraphics[width=\linewidth]{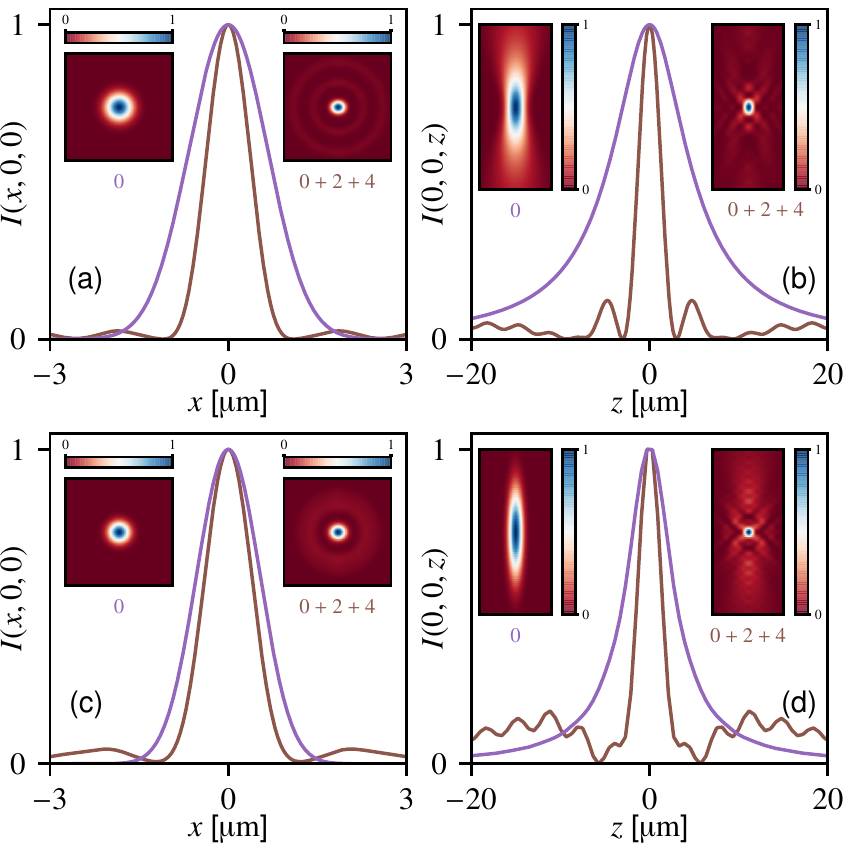}
\caption{Focused intensity distributions calculated within the vectorial Debye approximation for inputs \mbox{$\vec{E}_{0}$} (violet) and \mbox{$\vec{E}_{\Sigma}$} (brown). The numerical aperture is $\textrm{NA}=0.7$ and two filling factor values are compared: For $F_0=0.35$ (as in main article) (a) $x$-line cut transverse intensity profiles. The insets provide the x-y intensity distribution in the focal plane $z=0$. (b) $z$-line cut axial intensity profiles. The insets correspond to the x-z distribution. For $F_0=0.45$ (c,d). Plotted intensities for inputs \mbox{$\vec{E}_{0}$} (violet) and \mbox{$\vec{E}_{\Sigma}$} (brown) are normalized to their maximum values. Reproduced from Fig. 3 in the main text for convenience.}
\label{fig:SM1}
\end{figure}

\begin{figure}[htbp]
\centering
\includegraphics[width=\linewidth]{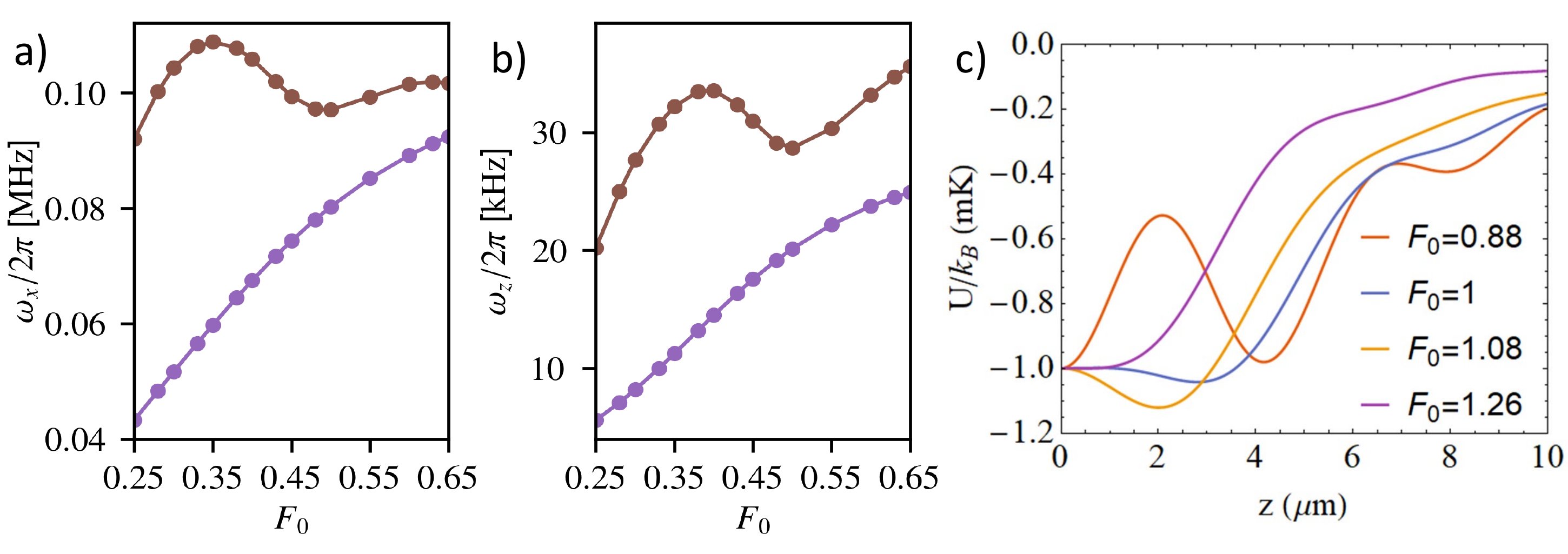}
\caption{Dependence of radial (x-cut, a) and axial (z-cut, b) trap frequencies ($\omega_x$ and $\omega_z$ at the bottom of the trap) as functions of the objective lens filling factor $F_0$, all for fixed numerical aperture $\textrm{NA} =0.7$. (Violet) the input field distribution is \mbox{$\vec{E}_{0}$}; (Brown) the input field distribution is \mbox{$\vec{E}_{\Sigma}$}. (c) Shows the evolution of the trap in the axial direction as the filling fraction increases. The reader will note that as $F_0$ changes from 1.0 to 1.26 the trap minimum lies away from $z=0$ and the trap frequency becomes ill-defined.}
\label{fig:SM2}
\end{figure}

With reference to Figure~\ref{fig:SM1}(a, b), we investigate trap volumes and frequencies beyond the paraxial using the vector theory with $F_0=0.35$, where the FWHMs for the `0+2+4' superposition inputs are $\Delta x_\Sigma = \SI{0.84}{\micro\meter}$,  $\Delta y_\Sigma = \SI{0.72}{\micro\meter}$ and $\Delta z_\Sigma = \SI{2.78}{\micro\meter}$. These parameters lead to a focal volume $V_{\Sigma} = \SI{1.7}{\cubic\micro\meter}$ for the central peak. For the $p=0$ input with $F_0=0.35$, the FWHMs of the central peaks for each direction are $\Delta x_0 = \SI{1.55}{\micro\meter}$, $\Delta y_0 =  \SI{1.51}{\micro\meter}$ and $\Delta z_0 = \SI{10.3}{\micro\meter}$, corresponding to a focal volume $V_{0} = \SI{24}{\cubic\micro\meter}$. The ratio of focal volumes defined via FWHMs for inputs with $p=0$ and the `0+2+4' superposition is then $V_{0}/V_{\Sigma} \simeq 14$. Moreover, for red-detuned optical traps associated with the line cuts in Figure~\ref{fig:SM1}(a, b) [Fig. 3(a, b) of the main text], we find trap frequencies for input \mbox{$\vec{E}_{\Sigma}$} to be $\omega_x^{\Sigma} =2\pi\times\SI{124}{\kilo\hertz}$ and $\omega_z^{\Sigma} = 2\pi\times\SI{33}{\kilo\hertz}$. The angular trap frequencies $\omega_x$, $\omega_y$ and $\omega_z$ are obtained by numerically solving the spatial time-independent Schr\"{o}dinger's equation for a single Cesium atom and, for simplicity here, a scalar optical trap potential with depth of \SI{1}{\milli\kelvin} for the $6\textrm{S}_{1/2}$ ground state of Cs. Trap frequencies as a function of the filling factor are shown in Fig. \ref{fig:SM2} (a-b) for both $E_0$ (purple curve) and $E_{\Sigma}$ (brown curve). Fig. \ref{fig:SM2}(c) shows how the trap evolves for changing filling factor and in particular explains the grey region in Fig. 4(b) of the main text.

\begin{figure}[htbp]
\centering
\includegraphics[width=\linewidth]{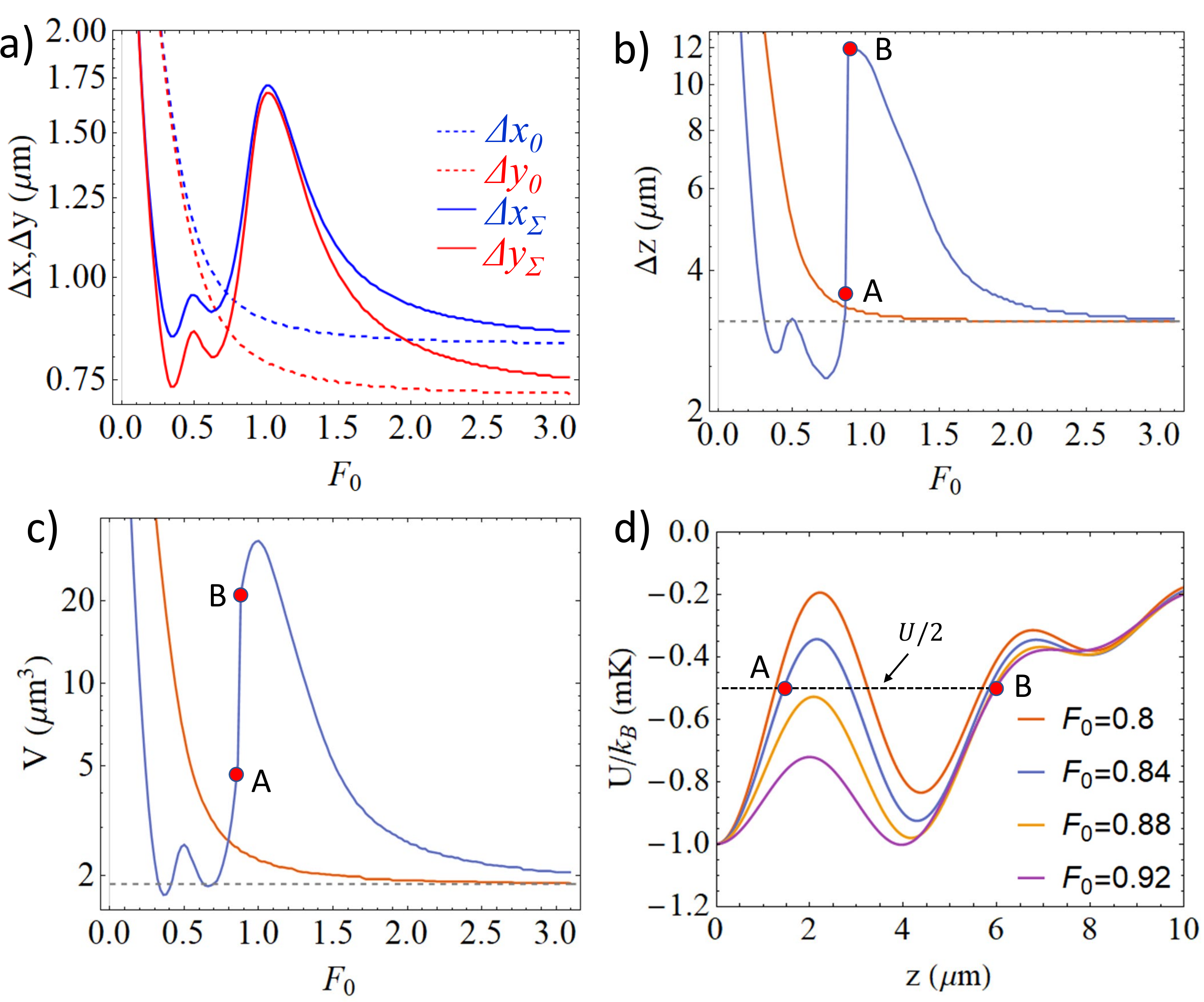}
\caption{Trap volume and dimensions as a function of the filling factor $F_0$. (a) Shows the dimensions in the $x$ and $y$ directions for a $p=0$ input field and our `0+2+4' input beam denoted with the $\Sigma$ subscript. (b) shows the $z$ (axial) direction character of the trap for different $F_0$ for $p=0$ (orange curve) and our $E_{\Sigma}$ beam (blue curve). Here we see a distinct decrease in trap size for $0.3<F_0<0.84$ and note that this lies below the lower limit of axial confinement for input $p=0$ beams, which approach the dotted line for large values of $F_0$. (c) The trap volume for $p=0$ (orange curve) and our $E_{\Sigma}$ beam (blue curve). Note the discontinuity in the blue curves in (b) and (c) from points A to B is a consequence of the merging center peak and side-lobes for  $0.88 \lesssim F_0 \lesssim 1$ which leads to a jump in the position of the FWHM along $z$. (d) illustrates how this aforementioned jump occurs, as $F_0$ increases the barrier between the main trap minimum at $z=0\mu$m and the side-lobe minimum at $z\simeq4.2 \mu$m drops below $U/2$ abruptly changing where the FWHM is defined from point A to point B.}
\label{fig:SM3}
\end{figure}

\begin{figure}[htbp]
\centering
\includegraphics[width=\linewidth]{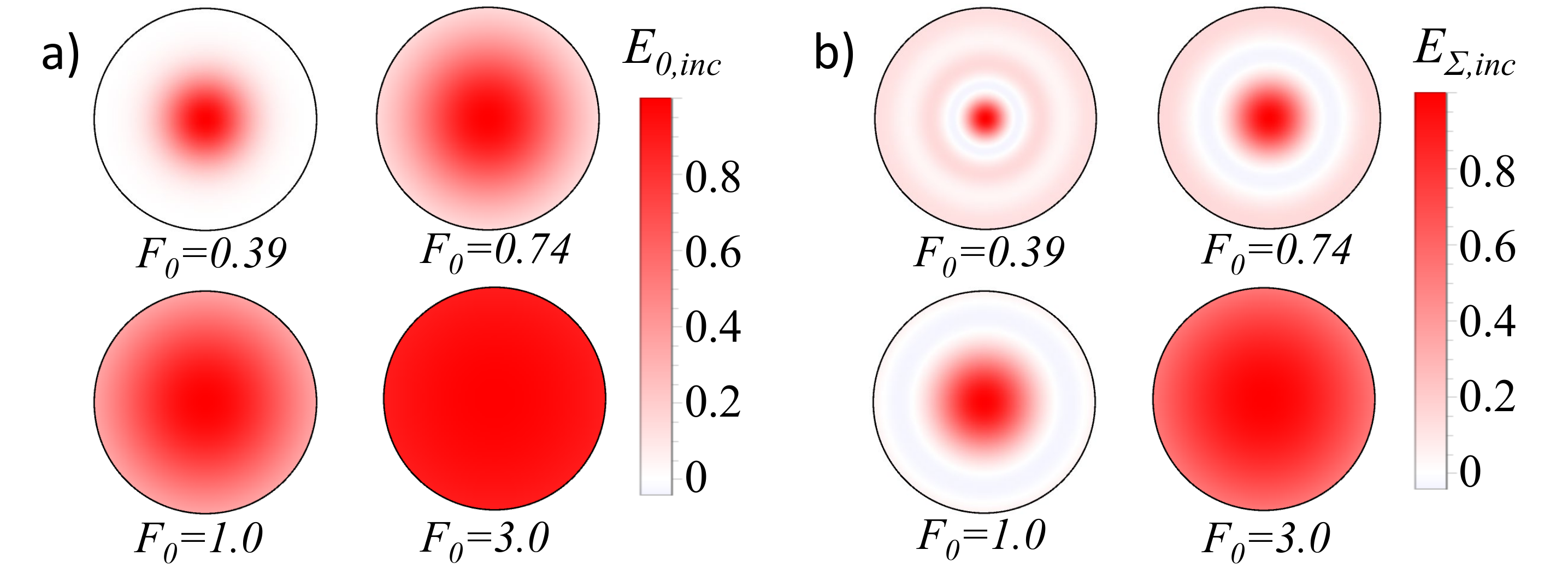}
\caption{For comparison, here are the input electric field profiles for the $p=0$ mode (a) and the `0+2+4' superposition (b) at filling factors $F_0=0.39$, $F_0=0.74$, $F_0=1.0$, and $F_0=3.0$ for the case of $\text{NA}=0.7$.}
\label{fig:SM4}
\end{figure}

Trap volume comparisons between the $p=0$ input field case and the `0+2+4' superposition input can be found in Fig. \ref{fig:SM3}. Here we show the reduction of trap volume of the $E_{\Sigma}$ input beam over the traditional $E_0$ Gaussian input which is evident for filling factors $0.3<F_0<0.84$ in Fig. \ref{fig:SM3} (b) and (c). The reader will note that the abrupt changes in the $E_{\Sigma}$ (blue curve) confinement and volume, which we have labeled `A' and `B', arise from our use of the FWHM to calculate volume. Using $U/2$ as the depth where we measure the trap width creates this change as the barrier between the intensity region at $z=0$ and the side-lobe at $z\simeq4\mu$m falls below $U/2$ and makes the FWHM point shift to point `A'. As a reference, Fig. \ref{fig:SM4} shows the field at the input aperture for increasing $F_0$ for both $E_0$ and $E_{\Sigma}$. Figure \ref{fig:SM4} shows the effect of increasing the filling factor $F_0$ for a finite size entrance aperture.

\section{Atom trajectory simulation}

In the Monte Carlo simulation of atom transport from free space tweezers to near surface traps, the atom sample is initialized from a sample of temperature $\SI{100}{\micro\kelvin}$ in $\SI{1}{\milli\kelvin}$ trap depth with position $\simeq \SI{600}{\micro\meter}$ away from the surface. The tweezer focus is first accelerated with acceleration $a=\SI{1}{\meter\per\square\second}$ towards the surface for $\SI{20}{\milli\second}$ and then moves at constant velocity for $\SI{10}{\milli\second}$ before decelerating with acceleration $a=\SI{-1}{\meter\per\square\second}$ to stop at the surface ($z= \SI{0}{\micro\meter}$) as shown in Fig.~\ref{fig:SM5}. To demonstrate the robustness of our scheme, we also simulated the atom transport with reflection coefficient $r=-0.3$ and $r=-0.8$ shown in the Main Text. Animations of typical atom trajectories are available in the accompanying supplementary files. 

\section{Media for 1D simulation}

\begin{figure}[ht]
\centering
\includegraphics[width=\linewidth]{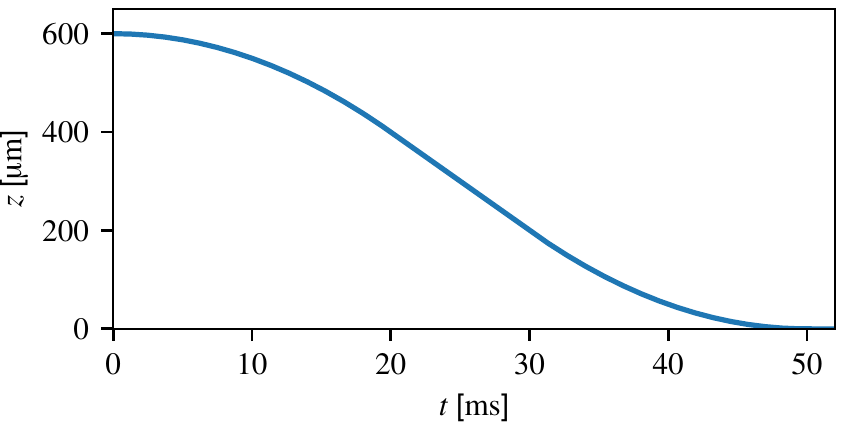}
\caption{Motion profile of the focal tweezer position used for the Monte Carlo simulations from Fig. 11 in the main text.}
\label{fig:SM5}
\end{figure}

Here we provide a description of the four movies associated with Fig. \ref{fig:SM6}, which are available at \url{http://dx.doi.org/10.22002/D1.1343}. All movies are generated under the paraxial approximation with \mbox{$w_0=\SI{1}{\micro\meter}$} and normalized to a trap depth of $U/k_B=\SI{1}{\milli\kelvin}$ in absence of the reflecting surface. The black dots represent individual, noninteracting atoms. The motion profile for the optical tweezer is given in Fig. \ref{fig:SM5}.

\section{Media for 3D simulation}

\begin{figure}[ht]
\centering
\includegraphics[width=0.96\linewidth]{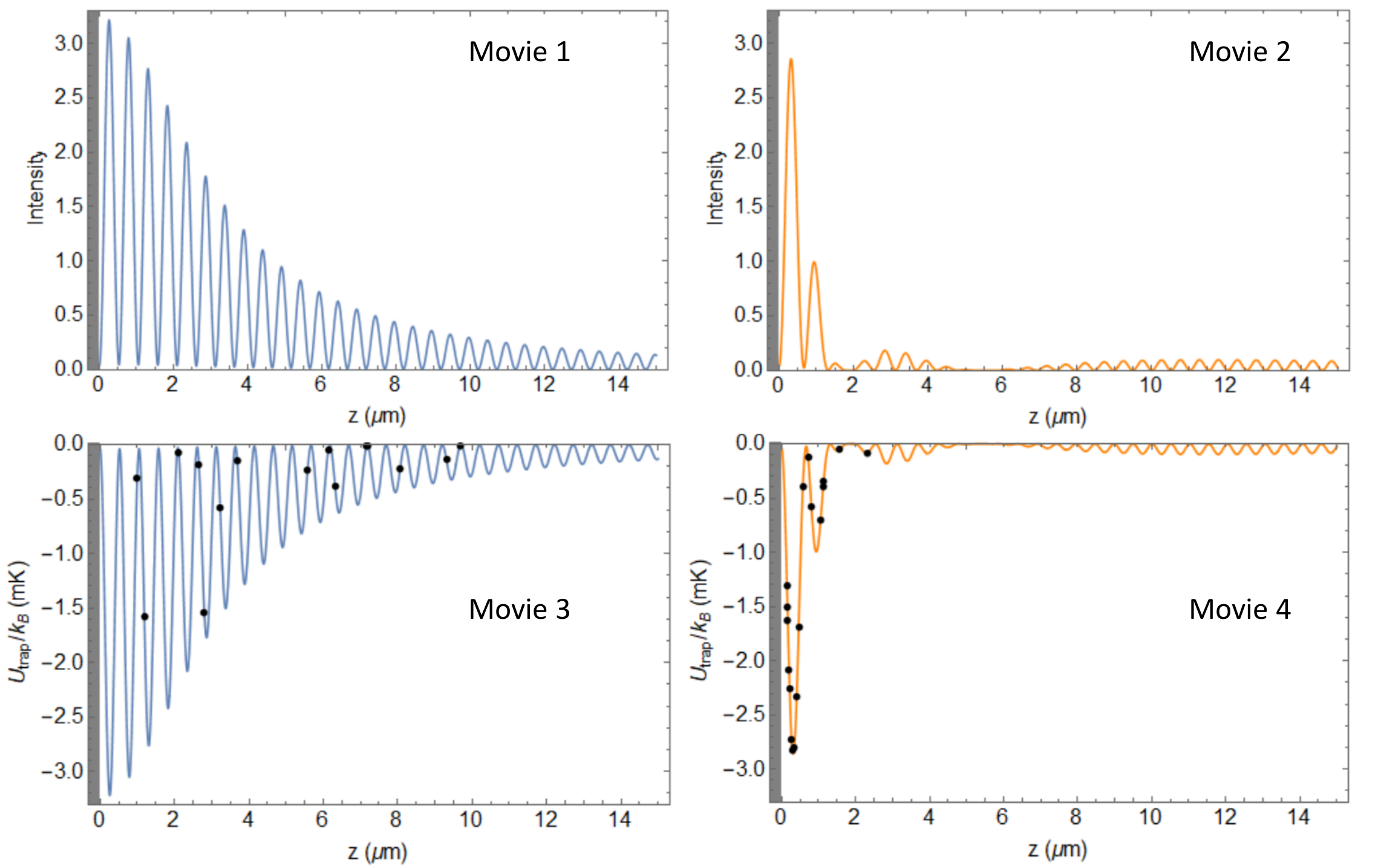}
\caption{Last frames of accompanying \mbox{Movies 1,2,3, and 4}, which are available at \url{http://dx.doi.org/10.22002/D1.1343}.  The simulations are carried out in the paraxial approximation as in Fig. 1 in the main text with \mbox{$w_0=\SI{1}{\micro\meter}$} and normalized to a trap depth of $U/k_B=\SI{1}{\milli\kelvin}$ in the absence of the reflecting surface at $z=0$, which has $r=-0.8$. Specifically, Movie 1 represents the intensity of \mbox{$\vec{E}_{0}$} tweezer with focus moving towards $z=0$; Movie 2 represents the intensity of \mbox{$\vec{E}_{\Sigma}$} tweezer with focus moving towards $z=0$; Movie 3 represents atom trajectory simulation for \mbox{$\vec{E}_{0}$} tweezer; and Movie 4 represents atom trajectory simulation for \mbox{$\vec{E}_{\Sigma}$}. The black dots represent the individual, noninteracting atoms.}
\label{fig:SM6}
\vspace{-4mm}
\end{figure}

Beyond simulations in 1D, we have also investigated atom transport in 3D for the moving tweezer potential $U(x(t),y(t),z(t))$, as shown in a movie at the following link \url{http://dx.doi.org/10.22002/D1.1346}. This animation shows the intensity of an \mbox{$\vec{E}_{\Sigma}$} tweezer with focus moving towards $z=0$. The black dots represent individual noninteracting atoms $\{i\}$ whose trajectories (i.e., positions $x_i(t),y_i(t),z_i(t)$) are driven by forces from $U(x_i(t),y_i(t),z_i(t))$. The parameters are as in Fig. \ref{fig:SM6}, again in the paraxial approximation with \mbox{$w_0=\SI{1}{\micro\meter}$} and normalized to a trap depth of $U/k_B=\SI{1}{\milli\kelvin}$ in absence of the reflecting surface.
The 3D results for trajectories are rendered into 2D for the animation by an orthographic projection into the $x,z$ plane.

\section{Preliminary results for generation of LG superpositions with an SLM}
\begin{figure*}[ht]
\centering
\includegraphics[width=\linewidth]{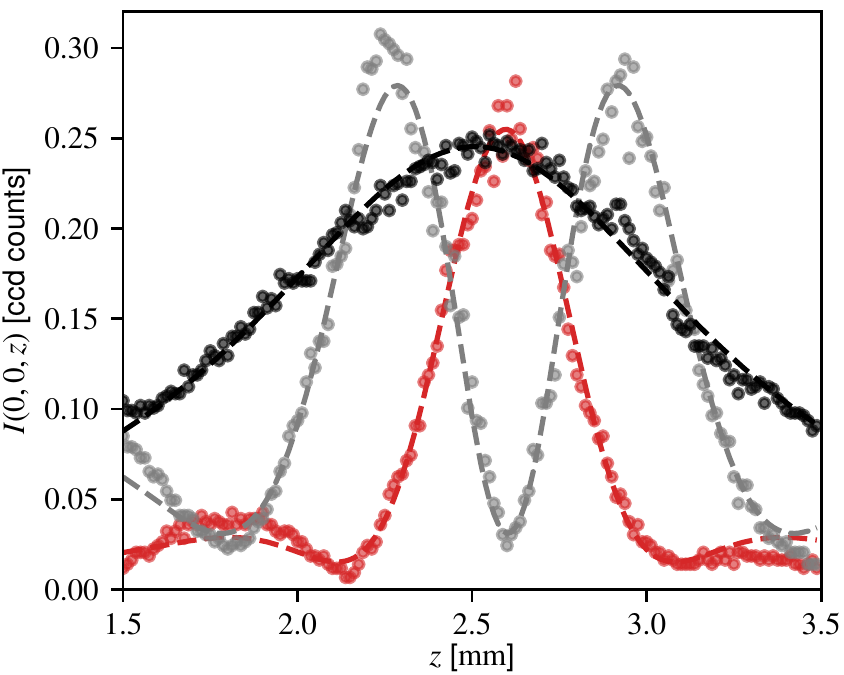}
\caption[Preliminary experimental result of LG mode superpositions.]{(points) Axial intensity profiles near the focus of a diffraction-limted aspheric lens (focal lens $f=\SI{50}{\milli\meter}$, NA$=0.2$) recorded directly (no imaging lens) with a CCD camera for three different input distributions: (black) single radial Laguerre-Gauss mode with $p=1$, (red) the coherent superposition $E_\Sigma$ and (gray) the coherent superposition of $E_{p=0}-E_{p=4}$. The corresponding dashed line are fits to the paraxial modes (plus constant background offset). The extracted focused Gaussian waist parameter is identical for all curves $w=\SI{15}{\micro\meter}$, in agreement with the expected value for the input waist $w_i = \SI{0.925}{\milli\meter}$ and wavelength $\lambda = \SI{852}{\nano\meter}$.}
\label{fig:SM7}
\end{figure*}

\begin{figure*}[ht]
\centering
\includegraphics[width=\linewidth]{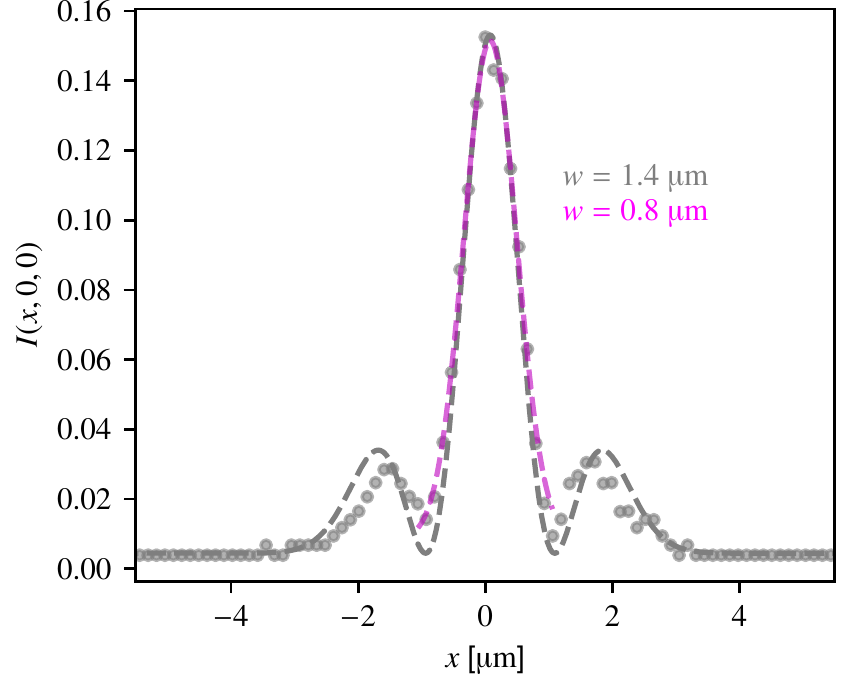}
\caption[Preliminary experimental result of LG mode superpositions.]{(points) Radial intensity profile of a single radial Laguerre-Gauss mode with $p=1$ measured at the focal plane of a microscope objective lens with NA=0.67 (along the input polarization direction). (gray dashed line) Fit with the paraxial mode profile with  extracted waist of $w=\SI{1.4}{\micro\meter}$. (magenta dashed line) Gaussian fit to the central lobe with the extracted waist of $w=\SI{0.8}{\micro\meter}$.}
\label{fig:SM8}
\end{figure*}

In Fig. \ref{fig:SM7} and Fig. \ref{fig:SM8} we show preliminary data for generation of LG superposition beams in the lab as in Fig. 2 of the main text. The SLM used here is the PLUTO-2-NIR-080 from Holoeye (\url{https://holoeye.com/}).

\newpage


\newpage

\newpage

\newpage

\newpage

\newpage

\newpage

\newpage

\newpage



\end{document}